# Approximation Algorithms for Semi-random Graph Partitioning Problems

Konstantin Makarychev  Yury Makarychev  Aravindan Vijayaraghavan
Microsoft Research  TTI Chicago  Princeton University

**Abstract**

In this paper, we propose and study a new semi-random model for graph partitioning problems. We believe that it captures many properties of real–world instances. The model is more flexible than the semi-random model of Feige and Kilian and planted random model of Bui, Chaudhuri, Leighton and Sipser.

We develop a general framework for solving semi-random instances and apply it to several problems of interest. We present constant factor bi-criteria approximation algorithms for semi-random instances of the Balanced Cut, Multicut, Min Uncut, Sparsest Cut and Small Set Expansion problems. We also show how to almost recover the optimal solution if the instance satisfies an additional expanding condition. Our algorithms work in a wider range of parameters than most algorithms for previously studied random and semi-random models.

Additionally, we study a new planted algebraic expander model and develop constant factor bi-criteria approximation algorithms for graph partitioning problems in this model.

## 1 Introduction

### 1.1 Overview

Graph partitioning problems are among the most fundamental problems in combinatorial optimization. They have numerous applications in science and engineering. They are also used as basic building blocks in many combinatorial algorithms. There has been extensive research on graph partitioning problems, which has been mostly focused on analyzing the worst case performance of optimization algorithms. Over the last two decades, poly-logarithmic approximation algorithms were developed for such fundamental graph partitioning problems as Minimum Bisection [34], Balanced Cut [30, 6], Multicut [21], Min Uncut [22, 1]. Yet, there has been little success in obtaining constant factor approximation algorithms for these problems, and some recent results [27, 28, 35, 36] suggest that this may even be hard, assuming the Unique Games conjecture [26] and its variants.

However, real-world instances of graph partitioning problems are very different from worst case instances. To take advantage of this, many attempts have been made [13, 17, 12, 25, 16, 19, 32, 10] to model average instances from practice and design algorithms that perform well in these models. The principal question now is — can we come up with a model, which on the one hand reasonably captures instances that often come up in practice, and on the other hand, leads to the development of new, interesting algorithms with good approximation guarantees in this model?

Moreover, if we were to believe that these basic graph partitioning problems were hard in worst-case it would be ideal to have a distribution of "hard" instances (as in random 3-SAT [18], planted clique [4], or densest $k$-subgraph [9]) that we can use as a test-bed for new algorithms. Further, in certain cases like the shortest vector problem on lattices [2, 3] and the densest $k$-subgraph problem [9], algorithms for an appropriate average-case distribution of instances have led to new insights for better algorithms in the worst case.

In this paper, we propose and study a new semi-random model for graph partitioning problems, which in our opinion captures many properties of real-world instances. We develop a general framework for analyzing semi-random instances of graph partitioning problems, and then present bi-criteria constant factor approximation algorithms for the "classical" problems of Balanced Cut, Sparsest Cut, Multicut and Min Uncut as well as for the Small Set Expansion problem (a problem which has recently attracted a lot of attention).

Before we proceed with the formal presentation of our model, let us discuss what we can reasonably assume about real-world instances. In a graph partitioning problem, the goal is to divide graph vertices into several parts, or clusters, so as to minimize the number of cut edges (subject to constraints that depend on a specific problem). When a practitioner solves a graph partitioning problem, she usually expects that the problem has a good solution — she believes that there is some underlying reason why there should be very few edges between clusters. That is, a real-world process that "generates" the graph instance adds an edge between clusters only when some random unexpected event happens. Therefore, in our opinion, it is reasonable to assume that edges between clusters are added at random. However, we cannot in general assume anything about edges within clusters (since their absence or presence does not affect the size of the cut between clusters). Additionally, in our model we assume that some random edges between cluster might be removed by the adversary (this assumption makes the model more robust). One could also view these edges between the clusters as random noise in an otherwise perfect clustering (partitioning).

This discussion leads to the following informal definition of semi-random instances: consider a set of vertices $V$ and some clustering of $V$. A semi-random graph $G$ on $V$ is a graph with *arbitrary (adversarial)* edges inside clusters and *random* edges between clusters (more generally, the set of edges between clusters might be a subset of a random set of edges).

Consider a toy example that illustrates why we believe that real-world instances are well described by our model. Suppose that we run a wiki website (or online store, online catalog etc). We track what pages our visitors read and construct a graph $G$ on the set of all wiki pages $V$ (see e.g., [33, 24]). If a visitor goes from page $A$ to page $B$, we connect $A$ and $B$ with an edge. What is the structure of this graph? We expect that a visitor will read one article, then read an article that explains some term mentioned in the first one, then read another article related to the second one and so on. Sometimes, of course, the visitor will move to a completely unrelated article on a different subject. Consequently, there will be two types of edges in our graph — edges between pages on the same subject, and edges between pages on different subjects. Edges of the first type are not random and show real connections between related articles. However, edges of the second type are essentially random. Say, an edge between articles "Ravioli" and "Register Allocation" is likely to be completely random and does not show any connection between articles; it just happened that the visitor first read an article about ravioli and then decided to read an article on register allocation; in contrast, an edge between articles "Ravioli" and "Dumplings" is not random and shows a real connection between these food items. To summarize, in our example

- edges between pages on one subject are not random (i.e. edges within a cluster);
- edges between pages on different subjects are random (i.e. edges between clusters).

So $G$ is a semi-random graph according to our model.

**Our semi-random model and results.** Now we are ready to give a formal definition. To be more specific, let us focus on the Balanced Cut problem.

**Definition 1.1.** (SEMI-RANDOM MODEL FOR BALANCED CUT) *We are given a set $V$ of $n$ vertices, and a parameter $\varepsilon$. In our model, a semi-random graph $G$ is generated as follows.*

1. *The adversary chooses a subset $S \subset V$ of $n/2$ vertices.*

2. *The nature chooses a set of random edges $E_R$ between $S$ and $V \setminus S$ and adds edges from $E_R$ to $G$. For every $u \in S$ and $v \in V \setminus S$, the edge $(u,v)$ belongs to $E_R$ with probability $\varepsilon$; choices for all edges $(u,v)$ are independent.*

3. *The adversary arbitrarily adds edges within $S$ and within $V \setminus S$.*

4. *The adversary deletes some edges between $S$ and $V \setminus S$.*

**Aim:** *The performance of the algorithm is measured by comparing the cost of edges cut to the expected number of edges in $E_R$ (the set of edges chosen at step 2).*

Note that the guarantees are not w.r.t the size of the cut $(S, V \setminus S)$ after step 4 or with the size of the optimal balanced cut. This is essential, since for example for $\varepsilon = 1$, $E_R = S \times (V \setminus S)$ the adversary can choose any graph $G$; so if we compared the cost of the cut with the cost of the optimal cut, our model would be the worst case model.

**Informal Theorem.** *Given a semi-random instance, our algorithm finds a balanced cut $(S', V \setminus S')$ with $|S'|, |V \setminus S'| = \Omega(n)$ of cost $O(|E_R|) = O(\varepsilon n^2)$ with high probability if*

$$\varepsilon > \sqrt{\log n}(\log \log n)^2/n.$$

**Informal Theorem.** *Given a semi-random instance, our algorithm finds a solution to the Small Set Expansion problem i.e., a subset $S \subset V$ of size $\rho n$, of cost $O(|E_R|) = O(\varepsilon \rho n^2)$ with high probability if*

$$\varepsilon \rho > \sqrt{\log n \log(1/\rho)}(\log \log n)^2/n.$$

Such results also hold for other basic graph partitioning problems like the Minimum Multicut, Sparsest Cut and Min Uncut (the complementary problem to MaxCut). The algorithm for the Small Set Expansion is not only interesting on its own, but can also be used to almost recover the original balanced cut under certain conditions. See Section 3 for a formal statement of the results. We remark that as $\varepsilon$ decreases, the problem becomes more challenging since the amount of randomness in the instances decreases.

Note that the algorithm does not necessarily find the planted cut $(S, V \setminus S)$ since in general this is impossible. Indeed the adversary can just delete all edges between $S$ and $V \setminus S$ and obtain an empty graph, or she can add every edge within $S$ and within $V \setminus S$ with probability $\varepsilon$ and obtain a random $G(n, \varepsilon)$ graph. In either case, our algorithm has no information about the planted cut $(S, V \setminus S)$.

However, if we assume that graphs induced by $S$ and by $V \setminus S$ are combinatorial expanders, we can almost recover sets $S$ and $V \setminus S$. This assumption is reminiscent of the stability assumption of Balcan, Blum and Gupta [7] for clustering problems (we need the extra condition on semi-randomness though). This assumption for planted partitioning problems can be justified in the implicit belief that approximately optimal solutions are close to the planted partition.

**Informal Theorem.** *There is a constant $C > 1$, such that for every constant $\eta > 0$, given a semi-random instance $G$ with combinatorial expansion $h(G[S]), h(G[V \setminus S]) \geq C\varepsilon n$ and $\varepsilon \eta > \sqrt{\log n \log(1/\eta)}(\log \log n)^2/n$, our algorithm finds with high probability the partition $(S, V \setminus S)$ up to an error of $\pm \eta n$ vertices.*

A similar result also holds for the Small Set Expansion problem.

**Planted Spectral Expander Model.** In this paper, we also develop bi-criteria approximation algorithms for graph partitioning problems on graphs with a planted spectral expander subgraph. Consider a graph $G$ with a (planted) balanced cut $(S, V \setminus S)$. Assume that the normalized algebraic expansion of the induced graph $G[S]$ is greater than the combinatorial expansion $h_{(S, V \setminus S)}$ of the cut by some constant factor. Then our algorithm finds a balanced cut with expansion $O(h_{(S, V \setminus S)})$. (Note that we do not impose any restrictions on the graph $G[V \setminus S]$ and on edges in the cut $(S, V \setminus S)$; this result also applies to the case when $G[S]$ is a random graph with the appropriate parameters.) We obtain a similar result for the Small Set Expansion problem. See Section 7 for details.

## 1.2 Prior Research

Our work extends prior research on random and semi-random instances of graph partitioning problems. The first random model, the planted random model, was introduced in 1984 by Bui, Chaudhuri, Leighton and Sipser [13]. In this model, we generate a graph on a set $V$ of size $n$ as follows. First, we randomly choose a subset $S$ of size $n/2$. Then we sample every edge between $S$ and $V \setminus S$ with probability $\varepsilon_1$, and every edge within $S$ and every edge within $V \setminus S$ with probability $\varepsilon_2 > \varepsilon_1$. Note that all choices in the planted random model are random (there are no adversarial choices), so the model describes a probability distribution on graphs. The model attracted a lot of attention and was studied in a series of papers by Dyer and Frieze [17], Boppana [12], Jerrum and Sorkin [25], Dimitriou and Impagliazzo [16], Condon and Karp [15] and Coja-Oghlan [14]. These papers explored several techniques for solving the problem — flow-based, combinatorial, spectral techniques, simulated annealing and go-with-the-leader technique. The algorithm of Boppana [12] finds the planted bisection $(S, V \setminus S)$ w.h.p. if $\varepsilon_2 - \varepsilon_1 > C\sqrt{\varepsilon_2 \log n/n}$. Later McSherry [32] obtained similar results for a more general class of graph partitioning problems.

Coja-Oghlan [14] extended the result of Boppana to the case when $\varepsilon_2 - \varepsilon_1 > C(\frac{1}{n} + \sqrt{\varepsilon_2 \log(n\varepsilon_2)/n})$. Note that if $\varepsilon_2 - \varepsilon_1 = o(\sqrt{\varepsilon_2 \log n/n})$ then the random graph has exponentially many minimum bisections and the planted bisection is not a minimum bisection w.h.p. [14]. The algorithm of Coja-Oghlan finds a minimum bisection rather than the planted bisection w.h.p.

In 2000, Feige and Kilian [19] proposed a more flexible *semi-random* model. The model adds an extra post-processing step to the random planted model: after a random graph is generated, the adversary may delete edges between $S$ and $V \setminus S$ and add new edges within $S$ and within $V \setminus S$. Semi-random instances of Feige and Kilian can have much more structure than random planted instances. Therefore, the model arguably captures real–world instances much better than the random model. From an algorithmic point of view, an important difference is that algorithms for the semi-random model of Feige and Kilian cannot overly exploit statistical properties of random graphs. In particular, spectral algorithms do not work for this model. Feige and Kilian [19] developed an SDP algorithm that finds the planted bisection if $\varepsilon_2 - \varepsilon_1 > C\sqrt{\varepsilon_2 \log n/n}$ (matching the bound of Boppana [12]).

In our semi-random model, the adversary has more power than in the model of Feige and Kilian. As in their model, the adversary can remove edges between $S$ and $V \setminus S$ but additionally she has absolute control over induced graphs $G[S]$ and $G[V \setminus S]$ (whereas in the model of Feige and Kilian, she could only add extra edges to random $G(\frac{n}{2}, \varepsilon_2)$ subgraphs inside $G[S]$ and $G[V \setminus S]$).

Our algorithm for Balanced Cut, while designed for a more general model, also works in a wider range of parameters than the algorithms of Boppana, and Feige and Kilian (note that the objective of our algorithms is slightly different, particularly we do not aim to recover the original partition precisely). To compare the algorithms, let us assume that probabilities $\varepsilon_1$ and $\varepsilon_2$ are of the same order of magnitude, $\varepsilon_1 = \Theta(\varepsilon)$ and $\varepsilon_2 = \Theta(\varepsilon)$. While the algorithm of Bopanna[12] and Feige and Kilian [19] require that $\varepsilon > C \log n/n$, we require only that $\varepsilon > C\sqrt{\log n}(\log \log n)^2/n$. (The algorithm of Coja-Oghlan works in even wider range of parameters; in the planted random model, it finds an optimal bisection, different from the planted partition, when $\varepsilon > C/n$.)

**Other Related Research.** Previously semi-random models for other combinatorial problems were studied by Blum and Spencer [11], Feige and Kilian [19], Feige and Krauthgamer [20], and Kolla, Makarychev and Makarychev [29]. Recently, Balcan, Blum and Gupta [7] (for clustering) and Bilu and Linial [10] (for MaxCut) investigated another very interesting model for real–world instances. They suggested that real–world instances are stable — there is a unique optimal solution $\mathcal{S}$, and every solution that is far away from $\mathcal{S}$ is much more expensive then $\mathcal{S}$. Their results however are not comparable with results for the planted random model, semi-random model of Feige and Kilian, and our results.

Guruswami and Sinop [23] recently presented approximation algorithms for partitioning problems on graphs with good spectral expansion. Their results (as well as results based on Cheeger's inequality) are not applicable to semi-random instances of Balanced Cut. A semi-random adversary can choose $k$ vertices in $S$ and remove all (or almost all) edges incident to them from the graph, making the the first $k$ eigenvalues of the Laplacian equal to zero (or close to zero). Moreover, even if the adversary does not modify a random planted graph, spectral algorithms based on Cheeger's inequality give only a trivial bound: the second eigenvalue of the normalized Laplacian of a random planted graph is $\lambda_2 \approx \varepsilon_1/\varepsilon_2$. Therefore, a $\Theta(\lambda_2)$-approximation algorithm finds a cut of cost $\Theta(\varepsilon_2 n^2)$, which is far from the cost of the optimal cut (it is actually within a constant factor of the cost of the worst/typical balanced cut in $G$).

**Comparison of Techniques.** Our approach was influenced by a recent paper of Kolla, Makarychev and Makarychev [29] on semi-random instances of Unique Games. In particular, we use Crude SDPs and the cut–long–edges method that were introduced in [29]. However, from the technical standpoint, this paper and [29] are very different and our algorithms require several new ideas . At high level, the algorithm of [29] (for the random edges adversarial constraints model) in one step finds a set of edges (constraints) $E^-$ that contains almost all corrupted edges, and then processes $E^-$. This technique does not work with graph partitioning problems since we can find only a set $E^-$ that contains a constant fraction of random edges (edges from $E_R$) in one step. In this paper, we have to iteratively solve an SDP or C-SDP program, and remove "long edges" at different scales in order to find almost all random edges. Moreover, we cannot just use the technique of [29] at each iteration by a number of reasons. Firstly, the argument of [29] inherently works only at one length scale (a constant scale); if we apply it at the same scale over and over, we will not make any progress. Secondly, the algorithm of [29] needs the set of edges between $S$ and $V \setminus S$ to be random, but this set is no longer random after the first iteration (it depends on choices of the algorithm that in turn depend on $E_R$). Finally, in order to find almost all edges from $E_R$ we have to cut "long edges" at smaller and smaller scales $\delta_t$. Each time we have to charge the number of cut edges to the cost of the SDP solution. Then the cost $OPT/\delta_t$ incurred at iteration $t$ will grow as $\delta_t$ goes to 0 and the total cost will significantly exceed $OPT$.

In this paper, we develop a technique of hidden solution sparsification that overcomes these difficulties. We design a procedure that at each iteration divides the graph into several pieces so that (i) the cost of the SDP solution in one of them is much smaller than $OPT$ (the condition is more involved for the C-SDP solution), and roughly speaking (ii) all other pieces do not have to be further partitioned. Then we recursively apply the algorithm only to the first set. The hidden solution sparsification technique is the main technical and conceptual contribution of our paper. We briefly discuss it in Section 1.3 of the Introduction.

It seems that existing algorithms for random and semi-random instances of graph partitioning problems cannot be adapted to our semi-random model, since they make too strong assumptions about their input graphs, which are not true in our model. In particular, the algorithm of Feige and Kilian crucially uses that the cost of the optimal SDP solution for a semi-random instance of the Balanced Cut problem *exactly* equals the cost of the cut $(S, V \setminus S)$. In our model, this is not the case — the cost of the SDP solution can be two times smaller than the cost of the cut.

### 1.3 Our Techniques

In this paper, we develop a general framework for solving semi-random instances of graph partitioning problems.

Let us give a very brief and informal outline of our approach. The core of our algorithms is a *hidden solution sparsification* step (HSS). This step is the same in all our algorithms. Intuitively, the goal of this step is to find and remove almost all edges from $E_R$ (edges between clusters) by removing at most $O(OPT)$ edges. More specifically, the HSS step finds a set of edges $E^-$ and divides the graph $G - E^-$ into a set $M$ and a number of sets $Z_i$ such that:

1. The cost of the optimal solution for the sub-instance on $G[M] - E^-$ is at most $OPT/polylog(n)$.

2. Roughly speaking, each $Z_i$ does not have to be further partitioned. Formally, we call this condition

Φ-feasibility. Say, for the Balanced Cut problem, this condition means that each set $Z_i$ contains at most $cn$ vertices (for $c < 1$); for Multicut, it means that each $Z_i$ contains at most one terminal from each source terminal pair.

3. All edges between $M$ and $Z_i$ and between sets $Z_i$ lie in $E^-$.
4. There are "few" edges in $E^-$. For Balanced Cut and Multicut, we require that $|E^-| < O(OPT)$; for Small Set Expansion we have a more involved condition.

Then we run an existing $polylog(n)$-approximation algorithm for the sub-instance on the graph $G[M] - E^-$ (e.g. run the algorithm of Arora, Rao and Vazirani [6] for Balanced Cut). First condition guarantees, that the algorithm finds a partition $\{M_i\}$ of $M$ of cost $O(OPT)$. We consider the combined partition $\{M_i, Z_j\}$ of $V$. When we solve Balanced Cut or Multicut, the total number of edges cut by this partition is $O(OPT)$. We join together some sets in $\{M_i, Z_j\}$ and obtain a feasible solution of cost $O(OPT)$ (this step depends on the problem at hand). When we solve Small Set Expansion, we get a weaker guarantee on the set $E^-$, so the cost of $\{M_i, Z_j\}$ might be very high. We use an extra post-processing step to find a subset of edges in $E^-$ that we really need to cut.

**Hidden Solution Sparsification.** For simplicity, let us focus now on the Balanced Cut or Multicut problem. We find the partition $\{M, Z_i\}$ as follows. We start with the trivial partition $M = V$ and then iteratively cut sets $Z_i$ from $M$. Once we cut a set $Z_i$, we do not further subdivide it. We ensure that after $t$ rounds the cost of the optimal solution for the sub-instance on $G[M] - E^-$ is $O(OPT/2^t)$ and that properties (2)–(4) hold. Then after $O(\log \log n)$ iterations, we get the desired partitioning.

At iteration $t$, we solve the SDP relaxation for the problem on $G[M] - E^-$ and obtain an SDP solution $\varphi : M \to \mathbb{R}^n$ (the solution assigns vector $\varphi(u)$ to each vertex $u$). Since the cost of the optimal solution for $G[M] - E^-$ is $O(OPT/2^t)$, the cost of the SDP solution is also $O(OPT/2^t)$. The solution defines a metric $d(u, v) = \|\varphi(u) - \varphi(v)\|^2$ on the set $M$. We analyze the metric at scale $\delta_t = \delta_0/2^t$ (where $\delta_0 > 0$ is an absolute constant). For every vertex $u$, consider the set $B_u = \{v : d(u, v) \leq \delta_t\}$ of vertices at distance at most $\delta_t$ from $u$. Let us say that a vertex $u$ is $\delta_t$-*light* if $|B_u| < \delta_t^2 n$, and that $u$ is $\delta_t$-*heavy* if $|B_u| \geq \delta_t^2 n$. Denote the set of heavy vertices by $H$ and light vertices by $L$. Broadly speaking, we first use a procedure to remove the heavy vertices $H$ (and further process them to get Φ-feasible sets $Z_i$), while cutting only a few edges (these cut edges are added to $E^-$). In the remaining graph $G[M] - E^-$ all vertices are light. We show that in such a solution, at most an $O(\delta_t^2)$ fraction of edges from $E_R$ are shorter than $\delta_t/2$. Here we crucially use that $E_R$ is a random set of edges (and, thus, the graph $G = (V, E_R)$ is "geometrically expanding"). We cut all edges in $G[M] - E^-$ longer than $\delta_t/2$ and add them to $E^-$. In the obtained graph $G[M] - E^-$ all edges are shorter than $\delta_t/2$, hence it contains at most $O(\delta_t^2 OPT)$ edges from $E_R$. Thus in the next iteration, the cost of the optimal solution for the sub-instance on $G[M] - E^-$ is $O(\delta_{t+1}^2 OPT)$ (as we need).

The Heavy Vertex Removal procedure finds new sets $Z_i$ that cover all heavy vertices $H$ in several rounds. In each round, we define a few sets $Z_i$; each $Z_i$ contains a subset of heavy vertices together with their $r$-neighborhoods (where $r \in (\delta_t, 2\delta_t)$). We cut sets $Z_i$ away, add edges from $Z_i$ to the rest of the graph to $E^-$ and then process remaining heavy vertices. We ensure that all sets $Z_i$ have a small diameter and this implies that sets $Z_i$ are Φ-feasible. We cut sets $Z_i$ so that sets $Z_i$ cut in one round are far away from each other and the total number of rounds is small. This guarantees that the total number of cut edges by this procedure is small (here, we use that each set $Z_i$ contains a ball $B_u$ for some heavy vertex $u$, and hence $Z_i$ is not very small).

To upper bound the number of edges cut by removing edges longer than $\delta_t/2$, we observe that the SDP value in iteration $t$ is $O(\delta_t^2 OPT)$. The number of these cut edges is $O(\delta_t^2 OPT/\delta_t) = O(\delta_t OPT)$. Thus, the number of edges cut in all iterations is $O(\sum_i \delta_i OPT) = O(OPT)$.

The algorithm for Small Set Expansion (SSE) requires several new ingredients. The main problem is that we cannot use the SDP relaxation of Bansal, Feige, Krauthgamer, Makarychev, Nagarajan, Naor and Schwartz [8] since it may assign zero vectors to all vertices in $S$. Instead we use a "Crude SDP" (C-SDP) for the problem. C-SDPs were recently introduced by [29]. The C-SDP for the Small Set Expansion is *not* a relaxation for the problem; its objective value may be much larger than the value of the optimal integral solution (in particular, the value of a C-SDP can be large even if the cost of the optimal solution is 0). To solve a semi-random instance of the Small Set Expansion problem, we first apply the HSS step. However, the number of edges in $E^-$ is bounded in expectation by the cost of the C-SDP solution and may be much larger than the cost of the optimal solution. So our algorithm cannot afford to cut all these edges. Nevertheless, we prove that the number of edges in $E^-$ incident to the set $S$ ($S$ is the optimal solution, which is not known to the algorithm) is bounded by $O(OPT)$ (the total number of edges in $E^-$ can be much larger than $OPT$). Then we show how to find a good solution by combining the SDP based SSE algorithm [8] with a new LP algorithm.

**Solution Purification.** As mentioned earlier, we show that if we additionally assume that graphs $G[S]$ and $G[V \setminus S]$ are combinatorial expanders in the Balanced Cut or Small Set Expansion problem, then we can almost recover sets $S$ and $G \setminus S$ (see Theorem 6.2). We do that by first finding a good approximate solution using our algorithm for Balanced Cut (or Small Set Expansion) and then improving the solution by repeatedly solving (semi-random) instances of the Sparsest Cut problem to obtain successively finer approximations to the planted partition.

**Organization.** In Section 2, we present our semi-random model and give definitions, which we use throughout the paper. In Section 3, we state the Hidden Solution Sparsification theorem. We describe our approximation algorithms for Balanced Cut, Multi Cut, Small Set Expansion and Sparsest Cut, which rely on the HSS theorem, in Sections 3.1, 3.2, 3.3 and 3.4. Then in Section 4, we prove the HSS theorem for graphs with the geometric expansion property. In Section 5, we show that semi-random graph satisfy the geometric expansion property and thus conclude the proof of our main result. In Section 6, we show that we can almost recover the original partitioning if all parts are combinatorial expanders. Finally, in Section 7, we study the planted algebraic expander model.

## 2 Preliminaries

In this section, we define some notation that is convenient for working with partitioning problems. Throughout the paper, we let $n$ to be the number of vertices $n = |V|$ and $\mathcal{H} = \mathbb{R}^n$. We denote by $C$ a fixed constant, and this will usually correspond to the approximation ratio given by the algorithm. We do not make an attempt to optimize the constants in this version of the paper.

We use the following notation: the $\ell_2^2$–diameter of a set $Z \subset \mathcal{H}$ equals $\mathrm{diam}(Z) = \max\{\|\bar{u} - \bar{v}\|^2 : \bar{u}, \bar{v} \in Z\}$; the $\ell_2^2$–ball of radius $r$ around a set $Z \subset \mathcal{H}$ is defined as $\mathrm{Ball}(Z, r) = \{\bar{u} \in \mathcal{H} : \exists \bar{v} \in Z \text{ s.t } \|\bar{u} - \bar{v}\|^2 \leq r\}$. We let $\mathrm{Ball}(v, r) = \mathrm{Ball}(\{v\}, r)$.

### 2.1 Partitions

**Definition 2.1.** *Let $V$ be a set of vertices. We say that $\mathcal{P}$ is a partition of $V$ into disjoint sets or simply partition, if $V = \bigcup_{P \in \mathcal{P}} P$ and every two $P', P'' \in \mathcal{P}$ are disjoint. For every vertex $u \in V$, denote by $\mathcal{P}(u)$ the unique set $P \in \mathcal{P}$ containing $u$.*

*Denote by $I_S : V \to \{0, 1\}$ the indicator function of the set $S \subset V$:*

$$I_S(u) = \begin{cases} 1, & \text{if } u \in S; \\ 0, & \text{otherwise.} \end{cases}$$

**Definition 2.2.** *Let $G = (V, E)$ be a graph and $\mathcal{P}$ be a partition of $V$. Define the set of edges cut by the partition as follows*

$$\mathrm{cut}(\mathcal{P}, E) \equiv \{(u, v) \in E : \mathcal{P}(u) \neq \mathcal{P}(v)\}.$$

*The cost of the cut equals the size of the set* $\text{cut}(\mathcal{P}, E)$:

$$\text{cost}(\mathcal{P}, E) \equiv |\text{cut}(\mathcal{P}, E)|.$$

*The cost of the cut restricted to a subset* $\mathcal{O} \subseteq V$ *only considers those edges which are incident on* $\mathcal{O}$:

$$\text{cost}_{|\mathcal{O}}(\mathcal{P}, E) \equiv |\{(u,v) \in \text{cut}(\mathcal{P}, E) : u \in \mathcal{O} \text{ or } v \in \mathcal{O}\}|$$

$$\equiv \sum_{(u,v) \in \text{cut}(\mathcal{P}, E)} \max\{I_\mathcal{O}(u), I_\mathcal{O}(v)\}.$$

## 2.2 Partitioning Problems

We now define three of the graph partitioning problems that we study in this paper.

**Definition 2.3.** (BALANCED CUT) *Given a graph* $G = (V, E)$, *the aim is to find a partition* $\mathcal{P}(P_1, P_2)$ *of* $V$ *with* $|P_1| = |P_2| = n/2$ *which minimizes* $\text{cut}(\mathcal{P}, E)$.

*A constant factor approximation algorithm finds a partition* $\mathcal{P}'(P_1', P_2')$ *with* $|P_1'|, |P_2'| \leq \beta n/2$ *for some fixed constant* $1 \leq \beta < 2$, *such that* $\text{cost}(\mathcal{P}', E) \leq O(1) \text{cost}(\mathcal{P}^*, E)$, *where* $\mathcal{P}^*$ *is an optimal balanced cut*[1].

**Definition 2.4.** (SMALL SET EXPANSION) *Given a graph* $G = (V, E)$ *and a parameter* $\rho \in (0, 1/2]$, *the aim is to find a partition* $\mathcal{P}(P_1, P_2)$ *of* $V$ *with* $|P_1| = \rho n$ *that minimizes* $\text{cost}(\mathcal{P}, E)$. *We will also be concerned with constant factor approximations (defined like in Balanced Cut).*

**Definition 2.5.** (MULTICUT) *Given a graph* $G = (V, E)$ *and a set of terminal pairs* $\{(s_i, t_i)\}_{1 \leq i \leq r}$, *the aim is to find a partition* $\mathcal{P}$ *of* $V$ *that separates all terminal pairs* $s_i, t_i$ *(i.e., for all* $i$, $\mathcal{P}(s_i) \neq \mathcal{P}(t_i)$) *and minimizes* $\text{cost}(\mathcal{P}, E)$.

## 2.3 Semi-random Models

We formally define the first semi-random model.

**Definition 2.6.** *Consider a set of vertices* $V$ *and a partition of vertices into disjoint sets* $\mathcal{P}$. *Let* $E_K = \{(u,v) : \mathcal{P}(u) \neq \mathcal{P}(v)\}$ *be the set containing all vertex-pairs crossing partition boundaries. Let* $\widetilde{E_K} = \{(u,v) : \mathcal{P}(u) = \mathcal{P}(v)\}$ *be the set containing all vertex-pairs not crossing partition boundaries. (Thus,* $(V, E_K \cup \widetilde{E_K})$ *is the complete graph on* $V$.) *Consider a random subset of edges* $E_R$ *of the set* $E_K$: *each edge* $(u,v) \in E_K$ *belongs to* $E_R$ *with probability* $\varepsilon$ *and these choices are independent. We define a random set of graphs* $SR(\mathcal{P}, \varepsilon)$ *as follows:*

$$SR(\mathcal{P}, \varepsilon) = \{G = (V, E) : E \subseteq E_R \cup \widetilde{E_K}\}.$$

*The optimal cost of the semi-random partition is defined as*

$$\text{sr-cost}(\mathcal{P}, \varepsilon) = \mathbb{E}|E_R| = \varepsilon |E_K|.$$

---

[1]This is sometimes referred to as $O(1)$ pseudo-approximation [6].

## 2.4 Local SDP Relaxations, Heavy Vertices and $\Phi$–Feasible Sets

**Definition 2.7.** *Let $V$ be a set of vertices. In this paper, we say that a map $\varphi : V \to \mathcal{H}$ is an SDP solution if vectors in $\varphi(V)$ satisfy $\ell_2^2$–triangle inequalities: for every $u, v, w \in V$, $\|\varphi(u) - \varphi(v)\|^2 + \|\varphi(v) - \varphi(w)\|^2 \geq \|\varphi(u) - \varphi(w)\|^2$.*

For instance, solutions to the SDP relaxations in Appendix C satisfy the above definition. The SDP solution $\varphi$ defines a metric on the vertices $V$ (given by $\|\varphi(u) - \varphi(v)\|^2$ for $u, v \in V$). The cost of a solution $\varphi$ corresponds to the total length of edges according to metric given by $\varphi$.

**Definition 2.8.** *Let $G = (V, E)$ be a graph, $\mathcal{P}$ be a partition of $V$, and $\mathcal{O}$ be a subset of $V$. Define the cost of an SDP solution $\varphi : V \to \mathcal{H}$ to be*

$$\text{sdp-cost}(\varphi, E) \equiv \frac{1}{2} \sum_{(u,v) \in E} \|\varphi(u) - \varphi(v)\|^2,$$

*and the cost of the SDP solution restricted to the set $\mathcal{O}$ to be*

$$\text{sdp-cost}_{|\mathcal{O}}(\varphi, E) \equiv \frac{1}{2} \sum_{\substack{(u,v) \in E \\ u \in \mathcal{O} \text{ or } v \in \mathcal{O}}} \|\varphi(u) - \varphi(v)\|^2.$$

For any SDP relaxation, the cost of the optimum (minimum) SDP solution lower bounds the value of the best integral solution. However, this lower bound may not hold when restricted to a subset $\mathcal{O} \subseteq V$. This motivates the following definition:

**Definition 2.9.** *Let $V$ be a set of vertices, $\mathcal{P}$ be a partition of $V$ and $\mathcal{O} \subseteq V$. We say that a non-empty set of SDP solutions $\Phi$ is a $\mathcal{O}$–local relaxation of $\mathcal{P}$ if there exists a constant $C \geq 1$ such that for every graph $G = (V, E)$ on $V$ and for*

$$\varphi = \arg\min_{\varphi \in \Phi} \text{sdp-cost}(\varphi, E) \equiv \arg\min_{\varphi \in \Phi} \frac{1}{2} \sum_{(u,v) \in E} \|\varphi(u) - \varphi(v)\|^2,$$

*the following inequality holds*

$$\text{sdp-cost}_{|\mathcal{O}}(\varphi, E) \leq C \, \text{cost}_{|\mathcal{O}}(\mathcal{P}, E).$$

Note that an SDP relaxation of a problem is always a $V$-local SDP relaxation of the optimal integral solution.

**Definition 2.10.** *Let $V$ be a set of vertices and $\Phi \subset \{\varphi : V \to \mathcal{H}\}$ be a set of SDP solutions. We say that a subset $S \subset V$ is $\Phi$–feasible if there exists $\varphi^* \in \Phi$ such that for every $u, v \in S$,*

$$\|\varphi^*(u) - \varphi^*(v)\|^2 \leq \frac{1}{4}.$$

$\Phi$–feasibility captures sets that require no further processing, to belong to a solution. For example, $\Phi$-feasible sets correspond to small enough sets for the Balanced Cut or Small Set Expansion problems, and to sets which do not contain any terminal pairs for the Multicut problem.

An SDP solution $\varphi$ classifies the vertices into two types (heavy or light) depending on the number of vertices in their $\delta$–neighborhoods.

**Definition 2.11.** *Let $V$ be a set of $n$ vertices, and $M \subseteq V$. Consider an SDP solution $\varphi : V \to \mathcal{H}$. We say that a vertex $u \in M$ is $\delta$–heavy in $M$ if the $\ell_2^2$-ball of radius $\delta$ around $\varphi(u)$ contains at least $\delta^2 n$ vectors from $\varphi(M)$ i.e., $|\{v \in M : \varphi(v) \in \text{Ball}(\varphi(u), \delta)\}| \geq \delta^2 n$. We denote the set of all heavy vertices by $H_{\delta, \varphi}(M)$.*

The following property of semi-random instances is crucially used in our algorithms.

**Definition 2.12.** (GEOMETRIC EXPANSION) *A graph $G = (V, E)$ satisfies the geometric expansion property with cut value $X$ at scale $\delta$ if for every SDP solution $\varphi : V \to \mathcal{H}$ and every subset of vertices $M \subseteq V$ satisfying $H_{\delta,\varphi}(M) = \varnothing$,*

$$|\{(u,v) \in E \cap (M \times M) : \|\varphi(u) - \varphi(v)\|^2 \leq \delta/2\}| \leq 2\delta^2 X.$$

*A graph $G' = (V, E')$ satisfies the geometric expansion property with cut value $X$ up to scale $2^{-T}$ ($T \in \mathbb{N}$) if it satisfies the geometric expansion property for every $\delta \in \{2^{-t} : 1 \leq t \leq T\}$.*

We can slightly simplify the definition[2] above by requiring that $\varphi$ satisfies the condition $H_{\delta,\varphi}(V) = \varnothing$ and $M = V$. See Section 5 for details.

In section 5, we will see that in semi-random instances $SR(\mathcal{P}, \varepsilon)$, the graph consisting of the random edges $(V, E_R)$ is geometrically expanding w.h.p. for sufficiently large $\varepsilon$.

## 3 Hidden Solution Sparsification and Applications

In this section, we state the main technical result of the paper, and then show how it can be used to obtain constant factor approximation algorithms for the Balanced Cut, Multicut and Small Set Expansion problems in the semi-random model.

**Theorem 3.1.** (HIDDEN SOLUTION SPARSIFICATION) *There exists a polynomial-time randomized algorithm that given a graph $G = (V, E)$, a separation oracle for an $\mathcal{O}$–local SDP relaxation $\Phi$ of a partition $\mathcal{P}$ (note: the set $\mathcal{O} \subset V$ and partition $\mathcal{P}$ are "hidden" and are not known to the algorithm), and a parameter $D = 2^T$ ($T \in \mathbb{N}$, $T > 1$), partitions the set of vertices $V$ into a set $M$ and a collection of disjoint sets $\mathcal{Z}$*

$$V = M \cup \bigcup_{Z \in \mathcal{Z}} Z,$$

*and also partitions the set of edges into two disjoint sets $E^+$ and $E^-$*

$$E = E^+ \cup E^-$$

*such that*

- *all edges cut by the partition $V = M \cup \bigcup_{Z \in \mathcal{Z}} Z$ lie in $E^-$ (i.e., $\mathrm{cut}(\{M\} \cup \mathcal{Z}, E) \subset E^-$), or in other words,*

$$E^+ \subset M \times M \cup \bigcup_{Z \in \mathcal{Z}} Z \times Z;$$

- *if the graph $(V, \mathrm{cut}(\mathcal{P}, E))$ satisfies the geometric expansion property with cut value $X$ up to scale $1/\sqrt{D}$, then (the expectation is taken over random bits of the algorithm)*

$$\mathbb{E}[\mathrm{cost}_{|\mathcal{O} \cap M}(\mathcal{P}, E^+)] \leq C \, X/D; \tag{1}$$

  *and*

$$|\{(u,v) \in E^- : u \in \mathcal{O} \text{ or } v \in \mathcal{O}\}| \leq C \, X; \tag{2}$$

- *each $Z \in \mathcal{Z}$ is $\Phi$–feasible.*

We first show how to construct constant factor approximation algorithms for Balanced Cut, Multicut and Small Set Expansion using the theorem. We prover the theorem in Section 4.

---
[2] We note that every Ramanujan expander is geometrically expanding with some parameters. However, we omit the details here.

## 3.1 Balanced Cut

We show that there exists a constant factor bi-criteria approximation algorithm for the Balanced Cut problem in the semi-random model with $\varepsilon \geq \Omega(\sqrt{\log n}(\log \log n)^2/n)$.

**Theorem 3.2.** *There exists a randomized polynomial-time algorithm, a function $f : \mathbb{N} \to [0, 1]$ tending to 0 as $n \to \infty$, and absolute constants $C, C_{BC}$, such that for every set of vertices $V$ of size $n$ (for simplicity assume $n$ is even), every partition $\mathcal{P} = \{L, R\}$, $|L| = |R| = n/2$, and every $\varepsilon \in (0, 1)$ with probability $1 - f(n) = 1 - o(1)$ over random choice of $SR(\mathcal{P}, \varepsilon)$ the following statement holds: for every $G = (V, E) \in SR(\mathcal{P}, \varepsilon)$ the algorithm returns a balanced partition of $V$ into sets $L'$ and $R'$ with $|L'|, |R'| \geq n/C$ and expected cost of the cut at most:*

$$\mathbb{E}\big[\mathrm{cost}(\{L', R'\}, E) \mid SR(\mathcal{P}, \varepsilon)\big] \leq C_{BC} \max\{\text{sr-cost}(\mathcal{P}, \varepsilon), n\sqrt{\log n}(\log \log n)^2\}.$$

*Particularly, if $\varepsilon \geq \sqrt{\log n}(\log \log n)^2/n$, then*

$$\mathbb{E}\big[\mathrm{cost}(\{L', R'\}, E) \mid SR(\mathcal{P}, \varepsilon)\big] \leq C_{BC} \, \text{sr-cost}(\mathcal{P}, \varepsilon) = C_{BC} \varepsilon \frac{n^2}{4}.$$

We use the standard SDP relaxation for the Balanced Cut problem. The SDP has a unit vector $\bar{u}$ for every vertex $u \in V$. All vectors satisfy $\ell_2^2$ triangle inequalities: for all $u, v, w \in V$:

$$\|\bar{u} - \bar{v}\|^2 + \|\bar{v} - \bar{w}\|^2 \leq \|\bar{u} - \bar{w}\|^2.$$

Finally, all vectors satisfy the spreading constraint (below we count every pair as $(u, v)$ and $(v, u)$):

$$\sum_{u,v \in V} \|\bar{u} - \bar{v}\|^2 \geq \frac{n^2}{2}.$$

The objective function of the SDP equals $1/2 \; \text{sdp-cost}(u \mapsto \bar{u}, E)$. (For clarity, we give the SDP in Appendix C.)

The SDP relaxation defines a set of feasible solutions $\Phi$. This set is a $V$-local SDP relaxation for $\mathcal{P}$, since every SDP relaxation is always a $V$-local SDP relaxation. Indeed, $\text{sdp-cost}_{|V}(\varphi, E) = \text{sdp-cost}(\varphi, E)$ for every $\varphi$ (just by definition), and particularly, for $\varphi^* = \arg\min_{\varphi \in \Phi} \text{sdp-cost}(\varphi, E)$,

$$\text{sdp-cost}_{|V}(\varphi^*, E) \equiv \text{sdp-cost}(\varphi^*, E) = \min_{\varphi \in \Phi} \text{sdp-cost}(\varphi, E)$$

$$\leq 2\,\mathrm{cost}(\mathcal{P}, E) \equiv 2\,\mathrm{cost}_{|V}(\mathcal{P}, E).$$

(The factor of 2 appears because of a different normalization of the objective function.)

In the algorithm below, we use the ARV algorithm for finding a balanced cut (in the worst-case) of Arora, Rao, and Vazirani [6]. We denote the approximation factor of the algorithm by $D_{ARV} = O(\sqrt{\log n})$. For simplicity of exposition we assume that $D_{ARV}$ is a power of 4.

---

**Balanced Cut Algorithm in Semi-random Model**

**Input:** a graph $G = (V, E) \in SR(\mathcal{P}, \varepsilon)$
**Output:** a cut $(L', R')$, with $|L'|, |R'| \geq n/C$

- Run the Hidden Solution Sparsification Algorithm with a separation oracle for $\Phi$ and obtain a set $M \subset V$, a partition $\mathcal{Z}$ of $V \setminus M$ in disjoint $\Phi$–feasible sets and two disjoint sets of edges $E^+$ and $E^-$ (with parameter $D = D_{ARV}$).

- Run the ARV algorithm on the graph $G = (V, E^+)$, and obtain a balanced partition $(L', R')$;
- **return** $(L', R')$.

---

**Analysis.** We show that every set $Z$ in $\mathcal{Z}$ is balanced. Every set $Z \in \mathcal{Z}$ is $\Phi$–feasible, that is, for some $\varphi \in \Phi$, $\varphi(Z)$ has $\ell_2^2$ diameter at most $1/4$. Thus,

$$\begin{aligned}
\frac{1}{2} \sum_{u,v \in V} \|\varphi(u) - \varphi(v)\|^2 &\leq \frac{1}{2} \sum_{u,v \in V} \max_{u,v \in V}(\|\varphi(u) - \varphi(v)\|^2) - \\
&\quad - \frac{1}{2} \sum_{u,v \in Z} \big(\max_{u,v \in V}(\|\varphi(u) - \varphi(v)\|^2) - \max_{u,v \in Z}(\|\varphi(u) - \varphi(v)\|^2)\big) \\
&\leq n^2 - \frac{7}{8} |Z|^2.
\end{aligned}$$

By the SDP spreading constraint, the left hand side is greater than or equal to $n^2/2$, thus $|Z| \leq \sqrt{4/7}\, n \leq 4/5\, n$.

By the Structural Theorem 5.1, with probability $1 - f(n) = 1 - o(1)$ for every graph $G = (V, E) \in SR(\mathcal{P}, \varepsilon)$, the graph $(V, \mathrm{cut}(\mathcal{P}, E))$ is geometrically expanding with cut cost

$$X = C \max\{\text{sr-cost}(\mathcal{P}, \varepsilon), n\sqrt{\log n}(\log \log n)^2\}$$

up to scale $1/\sqrt{D_{ARV}}$. Thus, by Theorem 3.1,

$$\text{cost}_{|M}(\{L, R\}, E^+) \leq C\, X/D_{ARV}.$$

Hence, there are at most $C\, X/D_{ARV}$ edges in $E^+$ going from $L \cap M$ to $R \cap M$. Observe, that $|L \cap M| \leq |L| = n/2$ and $|R \cap M| \leq |R| = n/2$. Therefore, there are at most $C\, X/D_{ARV}$ edges in $E^+$ cut by the the partition

$$V = (M \cap L) \cup (M \cap R) \cup \bigcup_{Z \in \mathcal{Z}} Z$$

(the only edges cut are the edges between $M \cap L$ and $M \cap R$) and each of the sets in the partition has size at most $4/5\, n$. These sets can be grouped into two balanced sets $L^*$ and $R^*$ with $|L^*|, |R^*| \geq 1/5\, n$. The ARV algorithm finds a possibly different balanced cut $(L', R')$ (with slightly weaker bounds on $|L'|, |R'|$). The number of edges cut in $E^+$ is bounded (in expectation) by $D_{ARV} \times C\, X/D_{ARV} = C\, X$. The number of edges cut in $E^-$ is bounded by $|E^-| \leq C\, X$.

### 3.2 Min Multicut and Min Uncut

The algorithm for the Multicut problem is similar to the algorithm for Balanced Cut. We use the standard SDP relaxation for Multicut: the SDP has a unit vector $\bar{u}$ for every vertex $u$; vectors $\bar{s}_i$, $\bar{t}_i$ corresponding to source–sink pairs $s_i$, $t_i$ are orthogonal ($\langle \bar{s}_i, \bar{v}_i \rangle = 0$); all vectors satisfy the $\ell_2^2$ triangle inequality constraints (please see Appendix C for details). The key observation is that every $\Phi$–feasible set $Z \in \mathcal{Z}$ has a small diameter w.r.t. some SDP solution and thus may not contain a source–sink pair $s_i$, $t_i$. Finally, to find a solution in $M$ we use the algorithm of Garg, Vazirani, and Yannakakis [21].

We get a constant factor approximation algorithm for the Multicut problem in the semi-random model with $\varepsilon \geq \log n (\log \log n)^2/n$.

**Theorem 3.3.** *There exists a randomized polynomial-time algorithm, a function $f : \mathbb{N} \to [0, 1]$ tending to 0 as $n \to \infty$, and an absolute constant $C$, such that for every set of vertices $V$ of size $n$, every partition $\mathcal{P}$, and every $\varepsilon \in (0, 1)$ with probability $1 - f(n) = 1 - o(1)$ over random choice of $SR(\mathcal{P}, \varepsilon)$ the following statement holds: for every $G = (V, E) \in SR(\mathcal{P}, \varepsilon)$ and every set of demands $(s_i, t_i)$ (satisfying $\mathcal{P}(s_i) \neq \mathcal{P}(t_i)$), the algorithm returns a partition $\mathcal{P}'$ of $V$ separating the demands ($\mathcal{P}'(s_i) \neq \mathcal{P}'(t_i)$) with expected cost of the cut at most:*

$$\mathbb{E}\big[\mathrm{cost}(\mathcal{P}', E) \mid SR(\mathcal{P}, \varepsilon)\big] \leq C \max\{\text{sr-cost}(\mathcal{P},\ \varepsilon), n \log n (\log \log n)^2\}.$$

A similar statement holds for the Min Uncut problem if $\varepsilon \geq \sqrt{\log n}(\log \log n)^2/n$. A semi-random instance of Min Uncut is generated as follows: the adversary first chooses an arbitrary subset $S$ of vertices, then the nature connects each pair of vertices $(u, v) \in S \times S \cup (V \setminus S) \times (V \setminus S)$ with an edge with probability $\varepsilon$, finally the adversary adds arbitrary edges between $S$ and $V \setminus S$, and removes some random edges. The problem can be restated as a cut minimization problem that falls in our framework (see e.g. [1]). Our algorithm for Min Uncut first runs the Hidden Solution Sparsification algorithm and then uses the algorithm of Agarwal, Charikar, Makarychev, and Makarychev [1] for Min Uncut. We defer the details to the journal version of the paper.

### 3.3 Small Set Expansion

Our algorithm for the Small Set Expansion (SSE) problem is the most involved, and uses the full power of the Hidden Solution Sparsification theorem.

**Theorem 3.4.** *There exists a randomized polynomial-time algorithm, a function $f : \mathbb{N} \to [0, 1]$ tending to 0 as $n \to \infty$, and an absolute constant $C$, such that for every set of vertices $V$ of size $n$, every partition $\mathcal{P} = \{S, V \setminus S\}$, $|S| = \rho n$ (for $\rho \in (0, 1/2)$) and every $\varepsilon \in (0, 1)$ with probability $1 - f(n) = 1 - o(1)$ over random choice of $SR(\mathcal{P}, \varepsilon)$ the following statement holds: for every $G = (V, E) \in SR(\mathcal{P}, \varepsilon)$, the algorithm given $G$ and $\rho$, returns a partition $\mathcal{P}' = (S', V \setminus S')$ of $V$ such that $|S'| = \Theta(\rho n), |S'| \leq |V|/2$ with expected cost of the cut at most:*

$$\mathbb{E}\big[\mathrm{cost}(\mathcal{P}', E) \mid SR(\mathcal{P}, \varepsilon)\big] \leq C \max\{\text{sr-cost}(\mathcal{P},\ \varepsilon), n \sqrt{\log n \log(1/\rho)}(\log \log n)^2\}.$$

*Particularly, if $\varepsilon \rho \geq \sqrt{\log n \log(1/\rho)}(\log \log n)^2/n$, then*

$$\mathbb{E}\big[\mathrm{cost}(\{L, R\}, E) \mid SR(\mathcal{P}, \varepsilon)\big] \leq C \, \text{sr-cost}(\mathcal{P}, \varepsilon) = C\varepsilon\rho(1 - \rho)n^2.$$

*Moreover, instead of requiring that $G = (V, E) \in SR(\mathcal{P}, \varepsilon)$, it suffices that the graph $(V, \mathrm{cut}(\mathcal{P}, E))$ is geometrically expanding with cut cost*

$$X = C' \max\{\text{sr-cost}(\mathcal{P},\ \varepsilon), n\sqrt{\log n \log(1/\rho)}(\log \log n)^2\}$$

*(for some absolute constant $C'$) up to scale $s(n, \rho) = \Omega(\sqrt{\log n \log \frac{1}{\rho}})$.*

By Theorem 5.1, for every graph $G \in SR(\mathcal{P}, \varepsilon)$, the graph $(V, \mathrm{cut}(\mathcal{P}, E))$ is geometrically expanding with cut cost

$$X = C' \max\{\text{sr-cost}(\mathcal{P},\ \varepsilon), n\sqrt{\log n \log(1/\rho)}(\log \log n)^2\}$$

up to scale $s(n, \rho) = \Omega(\sqrt{\log n \log \frac{1}{\rho}})$ with probability $1 - o(1)$ over random choice of $SR(\mathcal{P}, \varepsilon)$. We assume that the graph $(V, \mathrm{cut}(\mathcal{P}, E))$ is geometrically expanding. Otherwise, the algorithm fails (this happens with probability $o(1)$).

We use an analog of the Crude SDP (C–SDP) introduced in the paper of Kolla, Makarychev and Makarychev [29]. For each vertex $u \in V$ the C–SDP has a unit vector $\bar{u} \in \mathcal{H}$. All vectors satisfy triangle inequality constraints and spreading constraints (similar to constraints introduced in Bansal et al. [8]; note that their SDP would not work in our case: loosely speaking, it may "find" a good fractional cut that assigns zero vectors to the real solution): for every $u \in V$,

$$\sum_{v \in V} \langle u, v \rangle \leq \rho n.$$

We give the C-SDP in its entirety in Section C. Note that this SDP is not a relaxation for SSE. However, it turns out that this is a $S$–local SDP relaxation of partition $(S, V \setminus S)$ (see Lemma A.1). We now use the Hidden Solution Sparsification algorithm to find the set $M$ and a partition of $V \setminus M$ into $\Phi$–feasible sets $Z \in \mathcal{Z}$. Here $\Phi$ is the set of feasible C–SDP solutions. We set the weight of every vertex $u \in M$ to be the number of edges in $E^-$ incident on $u$: $w_u = |\{v : (u, v) \in E^-\}|$. $w_u$ corresponds to the cost we would pay for cutting the $w_u$ edges incident on $u$ from $E^-$, if $u$ were included in the solution (small set). Observe, that the weight of the "hidden" set $S$ is at most $C_1 X$ (for some absolute constant $C_1$, see (2)). Then, we consider two cases: $|M \cap S| \geq |S|/2$ and $|(V \setminus M) \cap S| \geq |S|/2$, depending on whether most of the hidden set $S$ vertices belong to $M$ or the pieces $Z \in \mathcal{Z}$ of the partition (the algorithm does not know which of the inequalities holds and tries both options).

**Case I:** This case is handled similar to the proof in Sections 3.1. Since most of $S$ (the hidden solution) belongs to $M$, we know that there is a good solution $M \cap S$ in $G(V, E^+)$ i.e. $S \cap M$ has size $\in [\rho n/2, \rho n]$ with weight $w(M \cap S) \leq C_1 X$, and there are at most $CX/D_{SSE}$ from $E^+$ going out of $S \cap M$. Now, we use the following theorem of Bansal et al. [8] which finds small non-expanding sets.

**Theorem 3.5.** (SPECIAL CASE OF THEOREM 2.1 [8], ARXIV VERSION) *There exists a polynomial-time algorithm ("SSE algorithm") that given as input a graph $G = (V, E)$, a set of positive weights $w_u$ ($u \in V$), $\rho \in (0, 1/2]$ and $W \in \mathbb{R}^+$, finds a non-empty set $S \subset V$ satisfying $|S| \in [\Omega(\rho n), 3\rho n/2]$, and $w(S) \equiv \sum_{u \in S} w_u \leq CW$, such that*

$$E(S, V \setminus S) \leq D_{SSE} \cdot \min\{E(S, V \setminus S) : |S| = \rho n, \ w(S) \leq W\},$$

*where $D_{SSE} = O(\sqrt{\log n \log(1/\rho)})$.*

**Remark:** This theorem is stated in a slightly different form in Bansal et al. [8]. We discuss the differences in Appendix (Section B).

We use this SSE algorithm on $G(V, E^+)$ to find a set $S'$ with $|S'| \in [\Omega(\rho n), 3\rho n/2]$, $w(S') \equiv \sum_{u \in S'} w_u \leq C \cdot C_1 X$, and

$$\text{cost}(\{S', V \setminus S'\}, E^+) \leq D_{SSE} \times CX/D_{SSE} \leq CX.$$

The total cost of the cut $E(S', V \setminus S')$ is bounded by the number of edges cut in $E^+$ and $E^-$ which is at most $CX$ and $CC_1 X$ respectively, which is $O(X)$ as needed.

If $\rho \in (1/3, 1/2)$, the set $S'$ may contain more than $n/2$ vertices, but no more than $3\rho n/2 \leq 3/4 \, n$. Then the algorithm returns $S'' = V \setminus S'$ satisfying $|S''| \in [n/4, n/2]$.

**Case II:** In this case, the hidden solution $S$ could mostly be spread arbitrarily among the pieces $Z \in \mathcal{Z}$ in $V \setminus M$. Here, our algorithm uses an LP to extract the solution from the set $V \setminus M$. The key observation is that this set is already partitioned into pieces of small size. Indeed, every $Z \in \mathcal{Z}$ is $\Phi$–feasible, and thus for some $\varphi \in \Phi$, $\text{diam}(\varphi(Z)) \leq 1/4$ and, consequently, for every $u, v \in Z$, $\langle \varphi(u), \varphi(v) \rangle = (\|\varphi(u)\|^2 + \|\varphi(v)\|^2 - \|\varphi(u) - \varphi(v)\|^2)/2 \geq 7/8$. Using the C-SDP spreading constraint (for an arbitrary $u \in Z$),

$$\sum_v \langle \varphi(u), \varphi(v) \rangle \leq \rho n,$$

we get $|Z| \leq 8/7 \, \rho n$.

The LP has a variable $x_v \in [0,1]$ for every vertex $v \in V \setminus M$; and the only constraint is that $\sum_{u \in V \setminus M} x_u \geq \rho n/2$). The objective function is

$$\min \sum_{u \in V \setminus M} w_u x_u + \sum_{\substack{(u,v) \in E^+ \\ u,v \in V \setminus M}} |x_u - x_v|. \tag{3}$$

The canonical solution to this LP is as follows: $x_u = 1$, if $u \in S \cap (V \setminus M)$; $x_u = 0$, otherwise. The LP cost of this solution is at most $CX$, because the first term in the objective function is bounded by $C_1 X$ (see (2)), the second term is bounded by the size of the $\text{cut}(\mathcal{P}, E)$, which is at most $C_2 X$ (The expected size of the cut equals $\text{sr-cost}(\mathcal{P}, \varepsilon)$; by the Chernoff bound the size of the cut is less than $2\,\text{sr-cost}(\mathcal{P}, \varepsilon)$ with very high probability). Thus, the cost of the optimal solution $\{x_u^*\}$, which we denote by $LP^*$, is at most $C\,X = (C_1 + C_2)X$. For an integral solution $S' \subset V$, we define the cost

$$f(S') = \sum_{u \in S'} w_u + |E^+(S', V \setminus S')\}| \tag{4}$$
$$\equiv \sum_{u \in S'} w_u + |\{(u,v) \in E^+ : u \in S', v \notin S'\}|.$$

For every $r \in [0,1]$ define $S_r = \{u : x_u^* \geq r\}$. The algorithm finds $r^*$ that minimizes the ratio $f(S_r)/|S_r|$ subject to $|S_r| \geq \rho n/4$ (note: $|S_1| = |V \setminus M| \geq \rho n/2$). Then it sorts all sets $Z \in \mathcal{Z}$ in order of increasing ratio $f(S_{r^*} \cap Z)/|S_{r^*} \cap Z|$ (ignoring empty sets) and gets a list $Z_1, \ldots Z_K$. It picks the first $k$ pieces such that

$$|Z_1 \cap S_{r^*}| + |Z_2 \cap S_{r^*}| + \cdots + |Z_k \cap S_{r^*}| \in [\rho n/4, 2\rho n],$$

and returns

$$S' = \bigcup_{i=1}^{k} Z_i \cap S_{r^*}.$$

Note, that such $k$ exists because each piece $Z_i \cap S_{r^*}$ has size at most $8/7 \, \rho n$ (as $|Z_i| \leq 8/7 \, \rho n$) and $\sum_{i=1}^{n} |Z_i \cap S_{r^*}| \equiv |S_{r^*}| \geq \rho n/4$.

**Analysis of Case II.** We first prove that

$$f(S_{r^*}) \leq 4LP^*/(\rho n) \cdot |S_{r^*}|.$$

Observe that

$$\int_0^1 f(S_r)dr = LP^* \qquad \int_0^1 |S_r|dr \geq \frac{\rho n}{2}.$$

The first equality easily follows from (3) and (4), the second equality follows from the LP constraint. Let $R = \{r : |S_r| \geq \rho n/4\}$. Then

$$\int_R |S_r|dr \geq \frac{\rho n}{2} - \int_{[0,1] \setminus R} |S_r|dr \geq \frac{\rho n}{4},$$

and, since $r^* = \min\{f(S_r)/|S_r| : r \in R\}$,

$$LP^* = \int_R f(S_r)dr \geq \int_R \frac{f(S_{r^*})}{|S_{r^*}|} |S_r|dr \geq \frac{f(S_{r^*})}{|S_{r^*}|} \cdot \frac{\rho n}{4}.$$

Thus, $f(S_{r^*}) \leq 4LP^*/(\rho n) \cdot |S_{r^*}|$. Using that edges in $E^+$ do not cross the boundaries of sets $Z_i$, we get

$$f(S_{r^*}) = \sum_{i=1}^{K} f(Z_i \cap S_{r^*}) \leq \frac{4LP^*}{\rho n} \sum_{i=1}^{K} |S_{r^*} \cap Z_i|.$$

Recall, that $\{f(Z_i \cap S_{r^*})/|S_{r^*} \cap Z_i|\}_i$ is an increasing sequence, thus

$$f(S') \equiv f(\bigcup_{i=1}^{k} Z_i \cap S_{r^*}) = \sum_{i=1}^{k} f(Z_i \cap S_{r^*})$$

$$\leq \frac{4LP^*}{\rho n} \sum_{i=1}^{k} |S_{r^*} \cap Z_i| = \frac{4LP^*}{\rho n} \cdot |S'| \leq 16LP^*.$$

## 3.4 Sparsest Cut

We now show how to find an approximate sparsest cut in a semi-random graph $G$ using the algorithm for Small Set Expansion. Specifically, we give an algorithm that for every subset $U \subseteq V$ intersecting each of the pieces of the planted partition $(S, T)$ (see below for details), returns a cut $(A, U \setminus A)$ of sparsity

$$\frac{E(A, U \setminus A)}{|A|} \leq O(\varepsilon n).$$

In Section 6, we show that the Sparsest Cut algorithm can be used to recover pieces $S$ and $T$ assuming that the graphs $G[S]$ and $G[T]$ have large expansion. We remark that while we are usually concerned with the case when $U = V$ for the sparsest cut problem, the following stronger statement is also useful for Section 6.

**Theorem 3.6.** *There exists a randomized polynomial-time algorithm, a function $f : \mathbb{N} \to [0, 1]$ tending to 0 as $n \to \infty$, and an absolute constant $C$, such that for every set of vertices $V$ of size $n$, every partition $\mathcal{P} = \{S, V \setminus S\}$ and every $\varepsilon, \eta \in (0, 1)$ satisfying $\varepsilon \eta \geq \sqrt{\log n}(\log \log n)^2/n$ with probability $1 - f(n) = 1 - o(1)$ over random choice of $SR(\mathcal{P}, \varepsilon)$ the following statement holds:*

*for every $G = (V, E) \in SR(\mathcal{P}, \varepsilon)$, every $U \subseteq V$ such that $|U \cap S| \geq \eta n$ and $|U \cap T| \geq \eta n$, the algorithm given $G$, returns a partition $(A, U \setminus A)$ of $G[U]$ with $|A| \leq |U|/2$ such that with probability exponentially close to 1,*

$$\frac{|E(A, U \setminus A)|}{|A|} < C_{SC}\varepsilon n. \tag{5}$$

*Proof Sketch.* We first give a proof assuming $\varepsilon\eta \geq \sqrt{\log n \log(1/\eta)}(\log \log n)^2/n$.

Our algorithm guesses the size of $|S \cap U|$, computes the size of $|T \cap U| = |U| - |S \cap U|$. Then, it runs the Small Set Expansion algorithm on $G[U]$ with $\rho = \min(|S \cap U|, |T \cap U|)/|U|$, obtains a set $A$ ($|A| \leq |U \setminus A|$) of size $\Theta(\rho|U|)$ and returns the cut $(A, U \setminus A)$. We need to show that the size of the cut $(A, U \setminus A)$ is at most $O(\varepsilon\rho|U|n)$, so that the sparsity of the cut is then $O(\varepsilon n)$.

Let us explain why we can use the Small Set Expansion algorithm for the graph $G[U]$ and why the algorithm finds a cut of cost at most $O(\varepsilon \rho |U|n)$. By the structural theorem (Theorem 5.1 part II), with probability $1 - o(1)$, for every $U \subset V$, the graph $(U, E \cap (U \times U) \cap (S \times T))$ (i.e., the bipartite graph between pieces $U \cap S$ and $U \cap T$) is geometrically expanding up to scale $\sqrt{\log n \log(1/\eta)}$ with cut value

$$X = C \max\{\text{sr-cost}(\mathcal{P}_{|U}, \varepsilon), n\sqrt{\log n \log(1/\eta)}(\log \log n)^2\}$$
$$= C \max\{\varepsilon|S \cap U| \cdot |T \cap U|, n\sqrt{\log n \log(1/\eta)}(\log \log n)^2\}.$$

Below, we assume that the graph $(U, E \cap (U \times U) \cap (S \times T))$ is geometrically expanding; otherwise our algorithm fails (which happens with probability $o(1)$ over the choice of $SR(\mathcal{P}, \varepsilon)$). Write the lower bound on $\varepsilon\eta$ and a trivial inequality on $\varepsilon |S \cap U| \cdot |T \cap U|$:

$$\sqrt{\log n \log(1/\eta)}(\log \log n)^2 \leq \varepsilon\eta n \leq \varepsilon \min(|S \cap U|, |T \cap U|)n;$$
$$\varepsilon |S \cap U| \cdot |T \cap U| \leq \varepsilon \min(|S \cap U|, |T \cap U|)n.$$

Together these inequalities give us an upper bound on $X$:

$$X \leq C\varepsilon \min(|S \cap U|, |T \cap U|)n \leq C\varepsilon \rho |U| n.$$

By Theorem 3.3, the Small Set Expansion algorithm returns a cut of size $O(X)$ (Here we use that the graph $(U, E \cap (U \times U) \cap (S \times T))$ is geometrically expanding up to scale $\sqrt{\log n \log(1/\eta)} \geq \sqrt{\log n \log(1/\rho)}$).

We showed that the algorithm finds a cut of sparsity $\alpha = O(\varepsilon n)$ in expectation. By Markov's inequality, it finds a cut of sparsity at most $2\alpha$ with probability at least $1/2$. So by repeating the algorithm many times and then picking the best solution, we can get a solution of cost at most $2\alpha$ with probability exponentially close to 1.

Finally, let us briefly explain how to get rid of the $\sqrt{\log(1/\eta)}$ factor in the lower bound on $\varepsilon\eta$. Observe that it suffices for our algorithm to find a set $A$ of size $|A| \in [\Theta(\rho|U|), |U|/2]$ i.e., we do not need a bound $|A| \leq O(\rho|U|)$. So we slightly modify the Small Set Expansion algorithm so that it works for smaller $\varepsilon\eta$, but possibly returns $|A| \gg \rho|U|$. In Case I of the algorithm (see Theorem 3.4), we use Theorem 2.1 (part I) instead of Theorem 2.1 (part II) of Bansal et al. [8] with $\rho = 1/2$. This algorithm returns a sparse cut of size at most $\rho|U| = |U|/2$ of sparsity $O(\sqrt{\log n})OPT$ (where $OPT$ is the optimal sparsity of the cut). We repeatedly apply this algorithm and obtain disjoint sets $A_1, \ldots, A_T$. After we get a set $A_i$, we remove it from $U$. We stop when $|\cup A_t| \geq \rho|U|/4$. We let $A = \cup A_t$. It is not hard to show that the sparsity of $A$ is at most $O(\sqrt{\log n})OPT$ (where $OPT$ is the value of the sparsest cut in $(U, E^+)$) and $|A| \in [\Theta(\rho|U|), 1/2|U|]$. The proof is similar to the proof of Theorem 2.1 (part II) in [8]. We omit it in this version of the paper. □

## 4 Hidden Solution Sparsification Algorithm

We now present the Hidden Solution Sparsification Algorithm and prove Theorem 3.1. The algorithm runs in $O(\log \log n)$ phases. In each round, we first solve the SDP on the current instance. The heavy vertices w.r.t. to this vector solution are first processed and removed using the algorithm from Section 4.1. In the remaining graph, we remove (cut) long edges to further "sparsify" the hidden solution ($E_R$) and produce the instance for the next phase.

---

**Hidden Solution Sparsification Algorithm**

**Input:** a graph $G = (V, E)$ and a separation oracle for a set of SDP solutions $\Phi \subset \{V \to \mathcal{H}\}$.
**Output:** partitions $V = M \cup \bigcup_{Z \in \mathcal{Z}} Z$ and $E = E^+ \cup E^-$.

- Let $M_0 = V$, $\mathcal{Z}_0 = \varnothing$, $E_0^+ = E$, $E_0^- = \varnothing$, $T = \frac{1}{2}\log_2 D$, and $\delta_t = 2^{-t}$ for all $t = 1, \ldots, T$.
- for $t = 1, \ldots, T$ do

    A. *Solve the SDP for the remaining graph:* Find
    $$\varphi_t = \arg\min_{\varphi \in \Phi} \text{sdp-cost}(\varphi, E_{t-1}^+ \cap (M_{t-1} \times M_{t-1})).$$

B. *Remove $\delta_t$–heavy vertices:* run Heavy Vertices Removal Algorithm (described in Section 4.1) with parameters $V$, $M_{t-1}$, $\varphi_t$, and obtain a collection of $\Phi$–feasible sets $\Delta \mathcal{Z}_t$. Add edges in $E_{t-1}^+$ cut by $\Delta \mathcal{Z}_t$ to the set $\Delta E_t^-$. Let

$$\mathcal{Z}_t = \mathcal{Z}_{t-1} \cup \Delta \mathcal{Z}_t; \quad M_t = M_{t-1} \setminus \bigcup_{Z \in \Delta \mathcal{Z}_t} Z.$$

C. *Remove $\delta_t$–long edges from $E^+$:* Find

$$L_t = \{(u,v) \in E^+ : u,v \in M_t, \|\varphi_t(u) - \varphi_t(v)\|^2 \geq \delta_t\}.$$

Let

$$E_t^+ = E_{t-1}^+ \setminus (\Delta E_t^- \cup L_t); \quad E_t^- = E_{t-1}^- \cup (\Delta E_t^- \cup L_t).$$

- **return** $M = M_T$, $\mathcal{Z} = \mathcal{Z}_T$, $E^+ = E_T^+$, $E^- = E_T^-$.

---

*Proof of Theorem 3.1.* We analyze the algorithm given above. We note that the step *A* of finding $\varphi_k$ can be performed in polynomial-time using semidefinite programming; the step *B* is performed using the algorithm described in the next subsection.

At every iteration, the algorithm removes all edges crossing the partition $\Delta \mathcal{Z}_t$ from $E_t^+$ and adds them to $E_t^-$, hence the first item of Theorem 3.1 holds. The third item holds, because every set $Z \in \mathcal{Z}$ belongs to some $\Delta \mathcal{Z}_t$ and, thus by Lemma 4.1 (see below), $\mathrm{diam}(\varphi_t(Z)) \leq 1/4$.

We now show that the second item of Theorem 3.1 holds. We first prove that

$$\mathrm{cost}_{|M_t}(\mathcal{P}, E_t^+) \leq 2\,X \cdot \delta_t^2$$

for every $t \in \{0, \ldots, T\}$. The Heavy Vertices Removal Procedure returns set $M_t$ that does not contain any $\delta_t$–heavy vertices w.r.t. $\varphi_t$ i.e., $H_{\delta_t, \varphi_t}(M_t) = \varnothing$ (see Lemma 4.1). Using the geometric expansion property of the graph $(V, \mathrm{cut}(E, \mathcal{P}))$, we get $\left|\{(u,v) \in \mathrm{cut}(\mathcal{P}, E) \cap (M_t \times M_t) : \|\varphi_t(u) - \varphi_t(v)\|^2 \leq \delta_t/2\}\right| \leq 2\delta_t^2 X$. The algorithm removes all $\delta_t/2$–long edges at step *C*, thus the set $E_t^+ \cap (M_t \times M_t)$ contains only edges $(u,v)$ for which $\|\varphi_t(u) - \varphi_t(v)\|^2 \leq \delta_t/2$. Combining this observation with the previous inequality, and using that edges in $E_t^+$ do not cross the boundary of $M_t$, we get

$$\mathrm{cost}_{|M_t}(\mathcal{P}, E_t^+) = \left|\mathrm{cut}(E, \mathcal{P}) \cap E_t^+ \cap (M_t \times M_t)\right| \leq 2\delta_t^2 X. \tag{6}$$

For $t = T$, we get $\mathrm{cost}_{|M}(\mathcal{P}, E^+) \leq 2\,X/D$.

Finally, we estimate the size of the set $\{(u,v) \in E^- : u \in \mathcal{O} \text{ or } v \in \mathcal{O}\}$. To do so, we use that $\Phi$ is a $\mathcal{O}$–local relaxation of the partition $\mathcal{P}$. For graph $G = (V, E_{t-1}^+ \cap (M_{t-1} \times M_{t-1}))$, we obtain inequality

$$\begin{aligned}
\mathrm{sdp\text{-}cost}_{|\mathcal{O}}(\varphi_t, E_{t-1}^+ \cap (M_{t-1} \times M_{t-1})) &\leq C_1 \mathrm{cost}_{|\mathcal{O}}(\mathcal{P}, E_{t-1}^+ \cap (M_{t-1} \times M_{t-1})) \\
&= C_1 \mathrm{cost}_{|\mathcal{O} \cap M_{t-1}}(\mathcal{P}, E_{t-1}^+) \leq 2CX \cdot \delta_{t-1}^2 \\
&= 8C_1 \delta_t^2 X.
\end{aligned}$$

The second line of the inequality follows from (6).

Now, we bound the number of edges removed from $E_{t-1}^+ \cap \mathcal{O}$ and added to $E_t^- \cap \mathcal{O}$ in terms of "sdp-cost". At step $t$, we add two sets of edges to $E^-$: $\Delta E_t^-$ and $L_t$. Since all edges $(u, v)$ in $L_t$ are $\delta_t/2$–long (i.e., $\|\varphi(u) - \varphi(v)\|^2 \geq \delta_t/2$),

$$\text{sdp-cost}_{|\mathcal{O}}(\varphi_t, E_{t-1}^+ \cap (M_{t-1} \times M_{t-1})) \equiv \sum_{\substack{(u,v) \in E_{t-1}^+ \cap (M_{t-1} \times M_{t-1}) \\ (u,v) \in \mathcal{O} \times V}} \frac{\|\varphi(u) - \varphi(v)\|^2}{2} \geq \frac{|L_t \cap (\mathcal{O} \times V)| \cdot \delta_t/2}{2}.$$

Hence, $|L_t \cap (\mathcal{O} \times V)| \leq 32 C_1 \delta_t X$. The probability that the Heavy Vertices Removal Procedure separates two vertices $u$ and $v$ connected with an edges in $E_{t-1}^+$ is at most $C_2 \left(\delta_t^{-1} + \delta_t^{-2} \mathbb{E}|M_{t-1} \setminus M_t|/n\right) \cdot \|\varphi(u) - \varphi(v)\|^2$ (see Lemma 4.1). Thus, the expected total number of edges in the set $\Delta E_t^- \cap (\mathcal{O} \times V)$ is at most

$$C_2 \left(\delta_t^{-1} + \delta_t^{-2} \frac{\mathbb{E}|M_{t-1} \setminus M_t|}{n}\right) \cdot \text{sdp-cost}_{|\mathcal{O}}(\varphi_t, E_{t-1}^+ \cap (M_{t-1} \times M_{t-1})) \leq 8 C_1 C_2 \left(\delta_t + \frac{\mathbb{E}|M_{t-1} \setminus M_t|}{n}\right) X.$$

The total number of edges in $E^- \cap (\mathcal{O} \times V)$ is bounded by

$$\sum_{t=1}^{T} \left(32 C_1 \delta_t + 8 C_1 C_2 \delta_t + 8 C_1 C_2 \cdot \frac{\mathbb{E}|M_{t-1} \setminus M_t|}{n}\right) X \leq (32 C_1 + 8 C_1 C_2 + 8 C_1 C_2) X.$$

□

## 4.1 Heavy Vertices Removal Procedure

In this section, we describe the algorithm which deals with the heavy vertices in a vector solution. Note that in the intended vector solution for all the above problems, all vertices are heavy (the intended solution for Balanced Cut has $n/2$ vectors at a fixed unit vector $\bar{v}_0$ and the rest $n/2$ of them at $-\bar{v}_0$). This algorithm also shows how we can take advantage of vector solutions which look like the intended solution (say, roughly low-dimensional solutions), with many heavy vertices.

**Lemma 4.1.** *There exists a polynomial-time algorithm that given a set of vertices $V$, an SDP solution $\varphi : V \to \mathcal{H}$, a subset $M \subseteq V$, finds a set of vertices $M' \subset M$ and a partition of $M \setminus M'$ into disjoint sets $Z \in \Delta \mathcal{Z}$ such that*

- *the set $M'$ does not contain any $\delta$–heavy vertices ($H_{\delta,\varphi}(M') = \varnothing$) w.r.t. $\varphi$.*
- $\text{diam}(\varphi(Z)) \leq 1/4$ *for every $Z \in \Delta \mathcal{Z}$;*
- *for every two vertices $u^*$ and $v^*$, the probability that $u^*$ and $v^*$ are separated by the partition is bounded as follows:*

$$\Pr(\exists Z \in \Delta \mathcal{Z} \text{ s.t. } I_Z(u^*) \neq I_Z(v^*)) \leq C \left(\delta^{-1} + \delta^{-2} \frac{\mathbb{E}[|M \setminus M'|]}{n}\right) \|\varphi(u^*) - \varphi(v^*)\|^2.$$

We remark that some of the heavy vertices $H_{\delta,\varphi}(M)$ may belong to $M'$, but they are not heavy anymore (w.r.t $M'$).

*Proof.* We use the following algorithm. If $\delta \geq 1/32$, we run the algorithm with $\delta' = 1/32$.

---

**Heavy Vertices Removal Procedure**

**Input:** a set of vertices $V$, a subset $M \subseteq V$, an SDP solution $\varphi : V \to \mathcal{H}$, a parameter $\delta \in (0, 1/32]$;
**Output:** a set $M \subseteq V$, partition $V \setminus M = \bigcup_{Z \in \Delta \mathcal{Z}} Z$;

- while ($H_{\delta,\varphi}(M) \neq \varnothing$)

    - Connect heavy vertices in $M$ at $\ell_2^2$ distance at most $4\delta$ with an edge and denote the new set of edges by $A = \{(u,v) \in H_{\delta,\varphi}(M) \times H_{\delta,\varphi}(M) : \|\varphi(u) - \varphi(v)\|^2 \leq 4\delta\}$.
    - Break graph $(H_{\delta,\varphi}(M), A)$ into connected components.
    - Pick a random $r \in [\delta, 2\delta)$.
    - *Remove components of small diameter:* For each connected component $U$ with $\mathrm{diam}(\varphi(U)) \leq 1/8$, let
    $$B_U = \{v \in M : \exists u \in U \text{ s.t. } \|\varphi(u) - \varphi(v)\|^2 \leq r\}.$$
    Denote the set of all connected components of diameter at most $1/8$ by $\mathcal{U}$.
    - *Remove a maximal independent set:* In the remaining set $H_{\delta,\varphi}(M) \setminus \bigcup_{U \in \mathcal{U}} U$ find a maximal independent set[3] $S$. For each $u \in S$, let $B_u = \{v : \varphi(v) \in \mathrm{Ball}(u,r)\}$.
    - Remove sets $B_U$ and $B_u$ from $M$:
    $$M = M \setminus \Big( \bigcup_{U \in \mathcal{U}} B_U \cup \bigcup_{u \in S} B_u \Big);$$

- **return** $M' = M$.

**Analysis.** It is clear that the algorithm always terminates in polynomial-time (since at every step at least one vertex is removed). When the algorithm terminates $H_{\delta,\varphi}(M) = \varnothing$ by the condition of the "while" loop. Every set $\varphi(B_u)$ removed from $M$ and added to $\Delta\mathcal{Z}$ at one of the iterations is contained in a ball of radius at most $2\delta$; every set $\varphi(B_U)$ is contained in the $2\delta$–neighborhood of a set $\varphi(U)$ (for some $U \in \mathcal{U}$) whose diameter is at most $1/8$. Thus, the diameter of each $\varphi(B_u)$ and $\varphi(B_U)$ is at most $1/8 + 4\delta \leq 1/4$.

Verify the third item of Lemma 4.1. Fix two vertices $u^*$ and $v^*$; and consider one iteration of the algorithm. We may assume that the algorithm first picks the independent set $S$ and a collection of connected components $\mathcal{U}$, and only then chooses random $r \in [\delta, 2\delta)$. Observe, that the distance between (images of) any two vertices in $S$ is at least $4\delta$ (because $S$ is an independent set), the distance between every two sets in $\mathcal{U}$ is at least $4\delta$ (because every $U \in \mathcal{U}$ is a connected component), and the distance between every $U \in \mathcal{U}$ and $u \in S$ is at least $4\delta$ (again because $U$ is a connected component, and $u \notin U$). Thus, $\varphi(u^*)$ may belong to at most one $\mathrm{Ball}(U, 2\delta)$ or $\mathrm{Ball}(u, 2\delta)$. If $\varphi(u^*) \in \mathrm{Ball}(u, 2\delta)$, then
$$\Pr(\varphi(u^*) \in \mathrm{Ball}(u,r), \ \varphi(v^*) \notin \mathrm{Ball}(u,r)) \leq \delta^{-1}\|\varphi(u) - \varphi(v)\|^2.$$
Of course, if $\varphi(u^*) \notin \mathrm{Ball}(u, 2\delta)$, then $\Pr(\varphi(u^*) \in \mathrm{Ball}(u,r), \ \varphi(v^*) \notin \mathrm{Ball}(u,r)) \leq \Pr(\varphi(u^*) \in \mathrm{Ball}(u,r)) = 0$.

The same statements hold if we replace $u \in S$ with $U \in \mathcal{U}$. Thus, at one iteration, the probability that $u^*$ belongs to a removed ball but $v^*$ does not belong to the same ball is at most $\delta^{-1}\|\varphi(u) - \varphi(v)\|^2$. Denote by $T$ the number of iterations of the algorithm. Then, the probability that $u^*$ and $v^*$ are separated at one of the iterations is at most $2\delta^{-1}\mathbb{E}[T]\|\varphi(u) - \varphi(v)\|^2$.

We now prove that at every iteration but possibly the last, the algorithm removes at least $\delta n$ vertices from $M$. Thus, $\mathbb{E}[T] \leq 1 + \mathbb{E}|M' \setminus M|/(\delta n)$, and the third item of Lemma 4.1 follows. Observe, that if the independent set $S = \varnothing$, then the algorithm terminates. If $S \neq \varnothing$, there exists at least one connected component $L$ with $\mathrm{diam}(\varphi(L)) \geq 1/8$. The maximal independent set in $L$ must contain at least $\Omega(\delta^{-1})$ vertices, since for every edge $(u,v) \in A$, $\|\varphi(u) - \varphi(v)\|^2 \leq 4\delta$. Thus, $|S| \geq \Omega(\delta^{-1})$. Since each $u \in S$ is $\delta$–heavy and $r \geq \delta$, $|B_u| \geq \delta^2 n$. Hence (using the fact that sets $B_u$ are disjoint),
$$\Big| \bigcup_{u \in S} B_u \Big| = \sum_{u \in S} |B_u| \geq \delta n.$$

□

---

[3]This is done independently of the random variable $r$, e.g., using a deterministic greedy algorithm.

# 5 Structural Theorem

We now prove that semi-random graphs are geometrically expanding, namely we prove that with high probability for every semi-random graph $G = (V, E) \in SR(\mathcal{P}, \varepsilon)$ the graph $(V, \text{cut}(\mathcal{P}, E))$ is geometrically expanding.

**Theorem 5.1.** *I. There exists a function $f : \mathbb{N} \to [0, 1]$ satisfying $\lim_{n \to \infty} f(n) = 0$ such that for every set of vertices $V$ of size $n$, every partition $\mathcal{P}$, and every $\varepsilon \in (0, 1)$, $D = 2^T$ ($T \in \mathbb{N}$, $T > 1$) with probability $1 - f(n) = 1 - o(1)$ the random set $SR(\mathcal{P}, \varepsilon)$ satisfies the following property: for every graph $G = (V, E) \in SR(\mathcal{P}, \varepsilon)$, the graph $(V, \text{cut}(\mathcal{P}, E))$ is geometrically expanding with cut cost*

$$X = C \max\{\text{sr-cost}(\mathcal{P}, \varepsilon),\ nD(\log^2 D)\}$$

*up to scale $1/\sqrt{D}$.*

*II. Moreover, a slightly stronger statement holds. For every set of vertices $V$ of size $n$, every partition $\mathcal{P}$, and every $\varepsilon \in (0, 1)$, $D = 2^T$ ($T \in \mathbb{N}$, $T > 1$) with probability $1 - f(n) = 1 - o(1)$ the random set $SR(\mathcal{P}, \varepsilon)$ satisfies the following property: for every graph $G = (V, E) \in SR(\mathcal{P}, \varepsilon)$ and every $U \subset V$, the graph $(U, \text{cut}(\mathcal{P}, E) \cap (U \times U))$ is geometrically expanding with cut cost*

$$X = C \max\{\text{sr-cost}(\mathcal{P}_{|U}, \varepsilon),\ nD(\log^2 D)\}$$

*up to scale $1/\sqrt{D}$. Here $\mathcal{P}_{|U} = \{P \cap U : P \in \mathcal{P}\}$ denotes the restriction of the partition $\mathcal{P}$ to the subset $U$.*

We defined Geometric Expansion in Section 2. We now give a slightly different definition of Geometric Expansion which is equivalent to Definition 2.12, but is more convenient for proving Theorem 5.1.

**Definition 5.2.** (GEOMETRIC EXPANSION; SEE DEFINITION 2.12) *A graph $G = (V, E)$ satisfies the geometric expansion property with cut value $X$ at scale $\delta$ if for every SDP solution $\varphi : V \to \mathcal{H}$ satisfying $H_{\delta, \varphi}(V) = \varnothing$,*

$$|\{(u, v) \in E : \|\varphi(u) - \varphi(v)\|^2 \leq \delta/2\}| \leq 2\delta^2 X.$$

*A graph $G = (V, E)$ satisfies the geometric expansion property with cut value $X$ up to scale $2^{-T}$ ($T \in \mathbb{N}$) if it satisfies the geometric expansion property for every $\delta \in \{2^{-t} : 1 \leq t \leq T\}$.*

**Claim 5.3.** *Definitions 2.12 and 5.2 are equivalent.*

*Proof Sketch.* It is easy to see that every graph satisfying Definition 2.12 satisfies Definition 5.2: we simply let $M = V$. Assume that $G = (V, E)$ satisfies Definition 5.2. Consider an SDP solution $\varphi : V \to \mathcal{H}$ and a set $M$ such that $\varphi_{\delta, \varphi}(M) = \varnothing$. Replace $\varphi$ with $\varphi'$: $\varphi'(u) = \varphi(u)$ if $u \in M$, and $\varphi'(u) = e_u$ otherwise, where $\{e_u\}_u$ is a collection of orthogonal unit vectors, orthogonal to all vectors $\varphi(u)$. The $\ell_2^2$–distance between every vector $\varphi'(u) = e_u$ ($u \in V \setminus M$) and any other vector $\varphi'(v)$ is at least 1. Thus, $H_{\delta, \varphi}(M) \subset H_{\delta, \varphi}(V) = \varnothing$. Hence,

$$\left|\{(u, v) \in E \cap (M \times M) : \|\varphi(u) - \varphi(v)\|^2 \leq \delta/2\}\right| =$$
$$= \left|\{(u, v) \in E : \|\varphi'(u) - \varphi'(v)\|^2 \leq \delta/2\}\right| \leq 2\delta^2 X.$$

□

*Proof of Theorem 5.1.* We use Definition 5.2 in this proof. Let $E_K = \{(u, v) \in V \times V : \mathcal{P}(u) \neq \mathcal{P}(v)\}$ and $E_R \subset E_K$ be the set of random edges chosen for the set $SR(\mathcal{P}, \varepsilon)$ as in Definition 2.6. Since $\text{cut}(\mathcal{P}, E) \subset E_R$ it suffices to show that the graph $(V, E_R)$ is geometrically expanding with high probability. We fix the parameter $\delta = 2^{-t}$ (where $1 \leq t \leq T$), and prove that the graph $(V, E_R)$ is geometrically expanding with cut value $X$ at scale $\delta$. Then we apply the union bound for all $T = \log_2 D$ possible choices of $\delta$.

We use the technique developed by Kolla, Makarychev and Makarychev [29]. Observe that the condition $H_{\delta,\varphi}(V) = \varnothing$ implies that
$$|\{v \in V : \|\varphi(u) - \varphi(v)\|^2 \leq \delta\}| \leq \delta^2 n,$$
and, consequently,
$$|\{(u,v) \in V \times V : \|\varphi(u) - \varphi(v)\|^2 \leq \delta\}| \leq \delta^2 n^2.$$

Thus we need to bound the probability of the bad event: *there exists an SDP solution $\varphi : V \to \mathcal{H}$ such that*
$$|\{(u,v) \in V \times V : \|\varphi(u) - \varphi(v)\|^2 \leq \delta\}| \leq \delta^2 n^2 \tag{7}$$
*and*
$$|\{(u,v) \in E_R : \|\varphi(u) - \varphi(v)\|^2 \leq \frac{\delta}{2}\}| \geq 2\delta^2 X. \tag{8}$$

We now show that if such $\varphi$ exists then there exists an embedding $\varphi' : V \to N_\delta$ to a relatively small set $N_\delta \subset \mathcal{H}$ satisfying slightly relaxed conditions:
$$|\{(u,v) \in V \times V : \|\varphi'(u) - \varphi'(v)\|^2 \leq \frac{3}{4}\delta\}| \leq \frac{5}{4}\delta^2 n^2, \tag{9}$$
and,
$$|\{(u,v) \in E_R : \|\varphi'(u) - \varphi'(v)\|^2 \leq \frac{3}{4}\delta\}| \geq \frac{3}{2}\delta^2 X. \tag{10}$$

Here $N_\delta \subset \mathcal{H}$ is a set of size $\exp(O(\log^2(1/\delta)))$ depending only on $\delta$. Then, we argue that such $\varphi'$ exists with very small probability.

**Claim 5.4.** *If $|E_R| \leq 2X$ and there exists $\varphi : V \to \mathcal{H}$ satisfying (7) and (8), then there exists $\varphi' : V \to N_\delta$ satisfying (9) and (10).*

*Proof.* We use the following simple lemma proved in [29].

**Lemma 5.5.** (LEMMA 3.7 [29], ARXIV VERSION) *For every positive $\zeta$, $\eta$ and $\nu$, there exists a set $N_\delta$ of unit vectors of size at most*
$$\exp\left(O(\zeta^{-2}\log(1/\eta)\log(1/\nu))\right)$$
*such that for every set of unit vectors $Z$ there exists a randomized mapping $\psi : Z \to N$ satisfying the following property: for every $u, v \in \mathcal{Z}$,*
$$\Pr((1+\zeta)^{-1}\|u-v\|^2 - \eta^2 \leq \|\psi(u) - \psi(v)\|^2 \leq (1+\zeta)\|u-v\|^2 + \eta^2) \geq 1 - \nu. \tag{11}$$

The proof of Lemma 5.5 is based on the Johnson–Lindenstrauss lemma: The set $N$ is an "epsilon–net" in a low dimensional space. To construct $\psi$ we first project $Z$ in a low dimensional space using the Johnson–Lindenstrauss transform and then "round" each vector to the closest vector in $N$. See [29] for details.

We set parameters $\zeta = 1/7$, $\eta^2 = \delta/8$ and $\nu = \delta^2/8$ and pick $N_\delta$ as in Lemma 5.5. Then we choose a deterministic $\psi(u) : \varphi(V) \to N$ such that the condition
$$\frac{7}{8}\|\varphi(u) - \varphi(v)\|^2 - \frac{\delta}{8} \leq \|\psi(\varphi(u)) - \psi(\varphi(v))\|^2$$
$$\leq \frac{8}{7}\|\varphi(u) - \varphi(v)\|^2 + \frac{\delta}{8}$$

holds for at least a $(1 - \delta^2/4)$ fraction of all pairs $u, v \in V$ and at least a $(1 - \delta^2/4)$ fraction of all edges $(u, v) \in E_R$ (the existence of such $\psi$ follows from (11), by the probabilistic method). Define $\varphi'(u) = \psi(\varphi(u))$. We get:

- for all but at most $\delta^2/4 \; n^2$ pairs $u,v \in V$, if $\|\varphi(u) - \varphi(v)\|^2 > \delta$, then $\|\varphi'(u) - \varphi'(v)\|^2 \geq 7/8 \; \delta - \delta/8 = 3/4 \; \delta$;

- for all but at most $\delta^2/4 \; |E_R| \leq \delta^2/2 \; X$ edges $(u,v) \in E_R$ if $\|\varphi(u) - \varphi(v)\|^2 \leq \delta/2$, then $\|\varphi'(u) - \varphi'(v)\|^2 \leq 8/7 \cdot \delta/2 + \delta/8 < 3/4 \; \delta$.

Therefore, inequalities (9) and (10) hold. □

Observe, that $\mathbb{E}|E_R| = \text{sr-cost}(\mathcal{P}, \varepsilon) \leq X$. Hence, by the Chernoff bound (for some absolute constant $C_1$),
$$\Pr(|E_R| \geq 2X) \leq e^{-C_1 X}.$$
Similarly, by the Chernoff bound, inequalities (9) and (10) simultaneously hold with probability at most $e^{-C_2 \delta^2 X}$. Thus, a fixed $\varphi' : V \to N$ satisfies (9) and (10) with probability (over random choice of $E_R$) at most $e^{-C_3 \delta^2 X}$. The total number of different embeddings $\varphi' : V \to N$ equals $|N|^n \leq \exp(C_4 n \log^2 D)$. By the union bound the probability that at least one such $\varphi'$ exists is at most $e^{-C_3 \delta^2 X + C_4 n \log^2 D} \leq e^{-n}$ here we use that $\delta^2 X \geq C n (\log^2 D)$ for sufficiently large $C$.

Part II follows from Part I by taking the union bound over all $2^n$ possible choices of the set $U$. We omit the details in this version of the paper. □

## 6 Recovering the Partitions in the Planted Model

In the case of the Balanced Cut and Small Set Expansion problems, we can obtain better guarantees when the sets of the partition $\mathcal{P} = \{S, T\}$ have enough expansion within them. Note that to recover the planted partition, we need some conditions on the graph expansion inside $G[S]$ and $G[T]$: otherwise, there may exist a sparse cut in $G$ cutting both $S$ and $T$ (for example, if the graphs $G[S]$ and $G[T]$ are random $G(n/2, \varepsilon)$ graphs, then the graph $G$ is a $G(n, \varepsilon)$, and thus the sets $S$ and $T$ are indistinguishable from other sets of size $n/2$). This assumption is in the flavor of planted instances of Balanced Cut (or Small Set Expansion problem), where the cut given by the partition $(S, T)$ is much sparser (sparser by a constant factor) than any cut inside the (adversarial) graph restricted to $S$ or $T$. (This assumption is also similar to the stability assumption of Balcan, Blum, and Gupta [7] for clustering problems, where a $c$-factor approximation to the partitioning problem is $\eta(c)$-close to the target partition.) In this case, we can find the partition $(S, T)$ up to $(1 + \eta)$-accuracy for some sub-constant $\eta > 0$ i.e., a partition differing from $(S, T)$ in at most $\eta n$ vertices. We obtain these guarantees by repeatedly defining instances of the Sparsest Cut problem, and using our algorithms for the semi-random model to obtain increasingly finer approximations to the planted partition.

**Definition 6.1.** *Denote by $h(G)$ the expansion of the graph $G = (V_G, E_G)$:*
$$h(G) \equiv \min_{\substack{S \subset V_G \\ 0 < |S| \leq 1/2 |V_G|}} \frac{E(S, V_G \setminus S)}{|S|}.$$

**Theorem 6.2.** *There exists a randomized polynomial-time algorithm, a function $f : \mathbb{N} \to [0,1]$ tending to 0 as $n \to \infty$, and positive absolute constants $C, C_{exp}$, such that for every set of vertices $V$ of size $n$, every partition $\mathcal{P} = \{S, T\}$, $|S| = \rho n$ (for $\rho \in (0, 1/2]$) and every $\varepsilon \in (0, 1)$, $\eta \in (0, 1)$, satisfying*
$$\eta \geq \frac{C\sqrt{\log n}(\log \log n)^2}{\varepsilon n},$$
*the following statement holds with probability $1 - f(n) = 1 - o(1)$ over a random choice of $SR(\mathcal{P}, \varepsilon)$: For every $G = (V, E) \in SR(\mathcal{P}, \varepsilon)$ satisfying $h(G[S]) \geq C_{exp} \varepsilon n$ and $h(G[T]) \geq C_{exp} \varepsilon n$, the algorithm given $G$ and $\varepsilon$, returns a partition $(X, Y)$ of $V$ such that*
$$|X \triangle S| = |Y \triangle T| \leq \eta n \quad \text{or} \quad |X \triangle T| = |Y \triangle S| \leq \eta n.$$

**Remark 1:** The conditions $h(G[S]) \geq C_{exp}\varepsilon n$ and $h(G[T]) \geq C_{exp}\varepsilon n$ can be slightly relaxed, by requiring that only sets of size at least $\eta n$ expand in $G[S]$ and $G[T]$.

**Remark 2:** We assume that $\eta \leq \rho/3$. Otherwise, if $\rho \leq \eta$, then the trivial solution $(\varnothing, V)$ satisfies the conditions of the theorem. If $\eta \in [\rho/3, \rho]$, we may replace $\eta$ with $\eta' = \rho/3$ and slightly change the absolute constant $C$.

Our algorithm relies on the Sparsest Cut algorithm for the semi-random model presented in Section 3.4. We denote the approximation factor of the Sparsest Cut algorithm by $C_{SC}$ (see 5). We let $C_{exp} = 4C_{SC}$. We will use this algorithm for finding approximate sparsest cuts in $G[X]$ for various $X \subset V$ satisfying $|X \cap S|, |X \cap T| \geq \eta n/2$ (sometimes these conditions on $X$ may be violated, then we assume that the algorithm returns a solution $A$, but the cut $(A, X \setminus A)$ may be arbitrarily bad). By Theorem 3.6, the Balanced Cut algorithm finds a cut of sparsity at most $C_{SC}\varepsilon n$ with probability exponentially close to 1 unless the graph $G$ does not satisfy the "strong geometric expansion" property described in Theorem 5.1, part II. This happens with probability $o(1)$; and in this case, the partition recovering algorithm described below fails as well.

We introduce a potential function $f$ that measures the quality of a partition $(X, Y)$:

$$f(X, Y) = C_{SC}\varepsilon n \min(|X|, |Y|) - |E(X, Y)|. \tag{12}$$

The algorithm presented below tries to maximize $f$ by finding non-expanding subsets $A$ in $X$ and moving them to $Y$ and finding non-expanding subsets $B$ in $Y$ and moving them to $X$.

**Algorithm.** The algorithm first finds an approximate sparsest cut $(X_0, Y_0)$ in $G$ using the Sparsest Cut algorithm for semi-random graphs. Then, it repeats the following refinement procedure: find approximate sparsest cuts $(A, X_t \setminus A)$ in the graph $G[X_t]$ and $(B, Y_t \setminus B)$ in the graph $G[Y_t]$ using the Sparsest Cut algorithm for semi-random graphs and

- if $f(X_t \setminus A, Y_t \cup A) \geq f(X_t, Y_t) + 1/4$, move $A$ from $X_t$ to $Y_t$ i.e., set $X_{t+1} = X_t \setminus A$ and $Y_{t+1} = Y_t \cup A$; otherwise,

- if $f(X_t \cup B, Y_t \setminus B) \geq f(X_t, Y_t) + 1/4$, move $B$ from $Y_t$ to $X$ i.e., set $X_{t+1} = X_t \cup B$ and $Y_{t+1} = Y_t \setminus B$.

The order in which the algorithm considers the cases above does not matter. After each iteration the algorithm increases the counter $t$. The algorithm stops and outputs the cut $(X_t, Y_t)$, when neither moving $A$ from $X$ to $Y$, nor moving $B$ from $Y$ to $X$ increases $f(X, Y)$ by at least $1/4$.

**Analysis.** Notice that the number of iterations of the algorithm is polynomial, since $f(X, Y)$ is upper bounded by $C_{SC}\varepsilon n^2$, lower bounded by $-|E|$, and at every iteration (but last) $f$ is increased by at least $1/4$. Thus, the algorithm runs in polynomial time. To prove that the algorithm works correctly, we need to show that the algorithm does not stop till $(X_t, Y_t)$ is $\eta$-close to the planted solution $(S, T)$ i.e., till $|X_t \triangle S| \leq \eta n$ or $|Y_t \triangle S| \leq \eta n$.

We first prove that $f(X_t, Y_t)$ is positive for every $t$. The Sparsest Cut algorithm finds a cut $(X_0, Y_0)$ of sparsity at most $C_{SC}\varepsilon n$, hence $f(X_0, Y_0) \equiv C_{SC}\varepsilon n \min(|X_0|, |Y_0|) - |E(X_0, Y_0)| > 0$. Since the sequence $f(X_t, Y_t)$ is increasing, $f(X_t, Y_t)$ is positive for every $t$. Consequently, the sparsity of every cut $(X_t, Y_t)$ is at most $C_{SC}\varepsilon n$.

We show that every relatively small set in $G$ expands.

**Claim 6.3.** *For every set $U \subset V$ of size at most $2\rho n/3$, $E(U, V \setminus U) \geq C_{exp}\varepsilon n|U|/2$.*

*Proof.* Since $h(G[S]) \geq C_{exp}\varepsilon n$, we have

$$E(U \cap S, S \setminus (U \cap S)) \geq C_{exp}\varepsilon n \cdot \min(|U \cap S|, |S \setminus (U \cap S)|)$$
$$\geq C_{exp}\varepsilon n|U \cap S|/2,$$

where the second inequality follows from $|U \cap S| \leq 2\rho n/3 \leq 2(|S| - |U \cap S|) = 2|S \setminus (U \cap S)|$. Similarly,

$$E(U \cap T, T \setminus (U \cap T)) \geq C_{exp}\varepsilon n |U \cap T|/2.$$

Thus, $E(U, V \setminus U) \geq C_{exp}\varepsilon n |U|/2$. $\square$

As a corollary, we get that $|X_t| \geq 2\rho n/3$ and $|Y_t| \geq 2\rho n/3$ for every $t$ (otherwise, the sparsity of the cut $(X_t, Y_t)$ would be large). To argue that the Sparsest Cut algorithm finds a $C_{SC}\varepsilon n$ sparse cut in $G[X_t]$ or $G[Y_t]$, we need to prove the following claim.

**Claim 6.4.** *Suppose that the partition $(X_t, Y_t)$ is not $\eta n$ close to the planted partition $(S, T)$ i.e., $|X_t \triangle S| = |Y_t \triangle T| > \eta n$ and $|X_t \triangle T| = |Y_t \triangle S| > \eta n$, then one of the following two statements holds:*

- $|X_t \cap S| \geq \eta n/2$ *and* $|X_t \cap T| \geq \eta n/2$*; or*
- $|Y_t \cap S| \geq \eta n/2$ *and* $|Y_t \cap T| \geq \eta n/2$.

*Proof.* The set $X_t$ is covered by $S$ and $T$, and thus $|X_t \cap S| \geq |X_t|/2$ or $|X_t \cap T| \geq |X_t|/2$. Assume that $|X_t \cap S| \geq |X_t|/2$. Then $|X_t \cap S| \geq |X_t|/2 \geq \rho n/3 \geq \eta n$. If also $|X_t \cap T| \geq \eta n/2$, we are done.

Otherwise, we have $|X_t \cap T| \leq \eta n/2$, and $|X_t \setminus S| = |X_t \cap T| \leq \eta n/2$. Consequently, $|Y_t \cap S| = |X_t \triangle S| - |X_t \setminus S| \geq \eta n - \eta n/2 = \eta n/2$. Also, $|Y_t \cap T| = |T| - |X_t \cap T| \geq \rho n - \eta n/2 \geq \eta n$.

The case $|T \cap X_t| \geq |X_t|/2$ is handled similarly. (Note that we have not used in the proof that $|S| \leq |T|$; we only used that $|T| \geq \rho n$.) $\square$

Apply Claim 6.4 and suppose without loss of generality that $|X_t \cap S| \geq \eta n/2$ and $|X_t \cap T| \geq \eta n/2$. Then, the set $X_t$ is partitioned in two pieces $X_t \cap S$ and $X_t \cap T$ each of size at least $\eta n/2$. Thus (as discussed in the beginning of the proof), the Balanced Cut algorithm finds a cut $(A, X_t \setminus A)$ (where $|A| \leq |X_t|/2$) of sparsity at most $C_{SC}\varepsilon n$. We now show that the cut $(A, V \setminus A)$ is large.

**Claim 6.5.** *Suppose that the graph $G$ is partitioned into three non-empty sets $U_1$, $U_2$, $U_3$, then one of the sets $U_i$ has large expansion: for some $i$,*

$$E(U_i, V \setminus U_i) \geq C_{exp}\varepsilon n |U_i|.$$

*Proof.* Observe that for one of the sets $U_i$, $|U_i \cap S| \leq |S|/2$ and $|U_i \cap T| \leq |T|/2$. For this set, $E(U_i \cap S, S \setminus U_i) \geq C_{exp}\varepsilon n |U_i \cap S|$ and $E(U_i \cap T, T \setminus U_i) \geq C_{exp}\varepsilon n |U_i \cap T|$. Hence, $E(U_i, V \setminus U_i) \geq C_{exp}\varepsilon n |U_i|$. $\square$

Consider the partition of $G$ into three sets $X_t \cap A$, $X_t \setminus A$ and $Y_t$. One of them has expansion $C_{exp}\varepsilon n$. It cannot be the set $Y_t$, since the expansion of $Y_t$ is at most $C_{SC}\varepsilon n$. Then,

$$\begin{aligned} E(X_t \setminus A, V \setminus (X_t \setminus A)) &\leq E(X_t, Y_t) + E(X_t \setminus A, A) \\ &\leq C_{SC}\varepsilon n |X_t| + C_{SC}\varepsilon n |A| \\ &\leq 3 C_{SC}\varepsilon n |X_t \setminus A| < C_{exp}\varepsilon n |X_t \setminus A|. \end{aligned}$$

Thus the set with expansion at least $C_{exp}\varepsilon n$ is $A$, that is, $E(A, V \setminus A) \geq C_{exp}\varepsilon |A| n$.

Estimate the change in the potential function $f$ after moving $A$ from $X_t$ to $Y_t$:

$$\begin{aligned} f(X_t \setminus A, Y_t \cup A) - f(X_t, Y_t) &\geq -C_{SC}|A|\varepsilon n - \big(E(A, X_t \setminus A) - E(A, Y_t)\big) \\ &= -C_{SC}|A|\varepsilon n - E(A, X_t \setminus A) + (E(A, V \setminus A) - E(A, X_t \setminus A)) \\ &= -C_{SC}|A|\varepsilon n - 2E(A, X_t \setminus A) + E(A, V \setminus A) \\ &\geq -C_{SC}|A|\varepsilon n - 2C_{SC}\varepsilon |A| n + 3/4\, C_{exp}\varepsilon |A| n + 1/4\, E(A, V \setminus A) \\ &= 1/4\, E(A, V \setminus A) \geq 1/4. \end{aligned}$$

# 7 Second Model: Algebraic Expansion inside Partitions

In the previous sections, we have seen that we can get much better approximation algorithms for partitioning problems when the edges $E_K$ crossing the boundaries of partition $\mathcal{P}$ satisfy some structural property (geometric expansion). In this section, we show that we can obtain good approximation algorithms for Balanced Cut and Small Set Expansion, when the edges $\widetilde{E_K}$ not crossing the partition boundaries satisfy some algebraic expansion condition. This is a much weaker condition than edges of $\widetilde{E_K}$ being chosen independently at random. More crucially, in this case, the edges $E_K$ can be arbitrary. Our algorithms are inspired by the results of [5, 31], where they infer global correlations between the vectors from local correlations and algebraic expansion.

**Theorem 7.1.** (BALANCED CUT) *There is a polynomial-time algorithm, that given a graph $G = (V, E)$ on $n$ vertices with a "planted" bisection $\mathcal{P} = \{P_1, P_2\}$ (not known to the algorithm) of cut value $\varepsilon m$, such that for some subset of edges $E_1 \subset E$ of size $|E_1| = m$, the graph $G_1 = (P_1, E_1)$ is a regular expander with a (normalized) algebraic expansion $\lambda(G_1) > 64\varepsilon$, finds a balanced cut of sparsity $O(\varepsilon)$.*

*Proof.* Consider the Balanced Cut SDP used in Section 3.1. Since this is a relaxation, the SDP value $SDP \leq \varepsilon m$. In particular,
$$\frac{1}{2}\text{sdp-cost}(u \to \bar{u}, E_1) \equiv \frac{1}{4}\sum_{(u,v)\in E_1}\|\bar{u}-\bar{v}\|^2 \leq \varepsilon m,$$
and, since $|E_1| = m$,
$$\frac{1}{4}\mathbb{E}_{(u,v)\in E_1}\left[\|\bar{u}-\bar{v}\|^2\right] = \frac{1}{4|E_1|}\sum_{(u,v)\in E_1}\|\bar{u}-\bar{v}\|^2 \leq \varepsilon.$$

For the regular graph $G_1$, we have
$$\lambda(G_1) \equiv \min_{\{\bar{u}\}_{u\in V}}\frac{\mathbb{E}_{(u,v)\in E_1}\left[\|\bar{u}-\bar{v}\|^2\right]}{\mathbb{E}_{u,v\in P_1}\left[\|\bar{u}-\bar{v}\|^2\right]},$$
thus
$$\frac{1}{4}\mathbb{E}_{u,v\in P_1}\left[\|\bar{u}-\bar{v}\|^2\right] \leq \frac{\varepsilon}{\lambda(G_1)} < \frac{1}{64}.$$

Hence, there exists $u^* \in P_1$ such that $\mathbb{E}_{v\in P_1}\left[\|\bar{u}^*-\bar{v}\|^2\right] \leq 1/16$. Denote $d(u,v) = \|\bar{u}-\bar{v}\|^2$. By Markov's inequality,
$$|\text{Ball}_d(u^*, \frac{1}{8}) \cap P_1| \geq \frac{|P_1|}{2} = \frac{n}{4}.$$

On the other hand, $|\text{Ball}_d(u^*, 1/4)| < 4/5\, n$ (as shown in Section 3.1).

We are ready to describe the algorithm: The algorithm guesses the vertex $u^*$ and picks a ball $S = \text{Ball}_d(u^*, r)$ of radius $r \in [1/16, 1/4]$ around $u^*$ with the smallest edge boundary. The cost of the cut $(S, V \setminus S)$ is at most $32 \cdot SDP$, and (since $\text{Ball}(u^*, 1/16) \subset S \subset \text{Ball}(u^*, 1/4)$),
$$\frac{n}{8} \leq |S| \leq \frac{4n}{5}.$$
$\square$

**Theorem 7.2** (Small Set Expansion). *There is a randomized polynomial-time algorithm, that given a graph $G = (V, E)$ with a "planted" partition $\mathcal{P} = \{P_1, P_2\}$ ($|P_1| = \rho n$) (not known to the algorithm) with $E(P_1, P_2) \leq \varepsilon m$ such that for some subset of edges $E_1 \subset V$ of size $|E_1| = m$, the graph $G_1 = (P_1, E_1)$ is a regular expander with a (normalized) algebraic expansion $\lambda(G_1) > 16\varepsilon$, finds a set $S$ of size $\rho n/4 \leq |S| \leq 2\rho n$ with expected cost of the cut $O(\varepsilon m)$.*

*Proof.* Let $\{\bar{u}\}_{u \in V(G)}$ be the solution of the C-SDP for the Small Set Expansion considered in Section 3.3. Denote

$$SDP_{|P_1} = \text{sdp-cost}(u \to \bar{u}, E_1) \equiv \frac{1}{2} \sum_{(u,v) \in E_1} \|\bar{u} - \bar{v}\|^2.$$

Since SDP is $P_1$-local relaxation of $\mathcal{P}$, $SDP_{|P_1} \leq OPT \equiv \varepsilon m$ (see Lemma A.1). We first proceed similarly to the proof of Theorem 7.1. Write,

$$\lambda \equiv \min_{\{\bar{u}\}_{u \in V}} \frac{\mathbb{E}_{(u,v) \in E_1}\left[\|\bar{u} - \bar{v}\|^2\right]}{\mathbb{E}_{u,v \in P_1}\left[\|\bar{u} - \bar{v}\|^2\right]},$$

then

$$\mathbb{E}_{u,v \in P_1}\left[\|\bar{u} - \bar{v}\|^2\right] \leq \frac{SDP_{|P_1}}{\lambda(G_1)} \leq \frac{OPT}{\lambda(G_1)} \leq \frac{1}{16}.$$

Hence, there is a vertex $u^* \in P_1$, such that

$$\mathbb{E}_{v \in P_1}\left[\|\bar{u} - \bar{v}\|^2\right] \leq 1/16.$$

Let $d(u,v) = \|\bar{u} - \bar{v}\|^2$. By the SDP spreading constraint (as shown in Section 3.3), $\text{Ball}_d(u^*, \frac{1}{4}) \leq 8/7 \, \rho n$.

Thus, for some radius $r \in [1/16, 1/4]$ (the algorithm can guess $u^*$ and $r$ by considering all possibilities), the set $S = \text{Ball}_d(u^*, r)$ contains at least $|P_1|/2$ vertices from $P_1$, but at most $8/7 \, \rho n$ vertices in total. Furthermore, the cost of the cut $E(P_1 \cap S, V \setminus S)$ is at most $32 OPT$.

The main difficulty and the main difference from the previous proof (Theorem 7.1) is that the set $S$ may contain vertices from $P_2$ and, moreover, it may cut many edges in $E(P_2 \cap S, V \setminus S)$. However, we have already dealt with a similar problem in Section 3.3. We use the LP (from the proof of Theorem 3.4, Case 2) to extract solution of cost at most $O(OPT)$ from $S$. The LP is feasible with LP value at most $O(OPT)$, because one integral "canonical" solution exists: it is the set $S' = P_1 \cap S$. Indeed, $E(P_1 \cap S, P_2 \cap S) \leq E(P_1, P_2) \equiv OPT$, and thus $E(P_1 \cap S, V \setminus (P_1 \cap S)) \leq 33 OPT$. □

## A   Local SDP Relaxation for SSE

**Lemma A.1.** *The set $\Phi$ of feasible solutions of the Crude SDP (C-SDP) given in Section 3.3 for the Small Set Expansion problem is a S-local relaxation of every partition $\mathcal{P} = \{S, V \setminus S\}$ (where $|S| = \rho n$).*

*Proof.* Let $\varphi = \arg\min_{\varphi \in \Phi} \text{sdp-cost}(\varphi, E)$. Denote $\bar{u} = \varphi(u)$. Define a new SDP solution

$$\bar{u}' = \begin{cases} \bar{e}_\perp & \text{if } u \in S \\ \bar{u} & \text{otherwise.} \end{cases}$$

where $\bar{e}_\perp$ is a unit vector orthogonal to all the vectors $\{\bar{v}\}_{v\in V(G)}$. This solution also satisfies the $\ell_2^2$–triangle inequalities, the spreading constraints (because $|S| \leq \rho n$ and for all $u \in S$, $v \in V \setminus S$, $\langle \bar{u}', \bar{v}' \rangle = 0 \leq \langle \bar{u}, \bar{v} \rangle$), and for all $u, v \in V$, $\langle \bar{u}', \bar{v}' \rangle \geq 0$. Thus, it lies in $\Phi$.

Compute the cost of the new solution and compare it with the cost of the optimal solution:

$$\text{sdp-cost}(u \to \bar{u}', E) = \frac{1}{2} \sum_{(u,v) \in E(G)} \|\bar{u}' - \bar{v}'\|^2$$

$$= \frac{1}{2} \sum_{\substack{(u,v) \in E(G) \\ u \in S \text{ or } v \in S}} \|\bar{u}' - \bar{v}'\|^2 + \frac{1}{2} \sum_{\substack{(u,v) \in E(G) \\ u, v \in V(G) \setminus S}} \|\bar{u}' - \bar{v}'\|^2$$

$$= \text{cost}_{|S}(\mathcal{P}, E) + \frac{1}{2} \sum_{\substack{(u,v) \in E(G) \\ u, v \in V(G) \setminus S}} \|\bar{u} - \bar{v}\|^2.$$

The cost $\text{sdp-cost}(u \to \bar{u}, E)$ of the optimal solution $\bar{u}$ equals

$$\text{sdp-cost}_{|S}(u \to \bar{u}, E) + \frac{1}{2} \sum_{\substack{(u,v) \in E(G) \\ u, v \in V(G) \setminus S}} \|\bar{u} - \bar{v}\|^2.$$

Thus, $\text{sdp-cost}_{|S}(u \to \bar{u}, E) \leq \text{cost}_{|S}(\mathcal{P}, E)$. $\square$

## B  Remark on Theorem 3.5

Theorem 3.5 is stated in a slightly different form in Bansal et al. [8]. We use Theorem 2.1 (part II) [8, p. 6; arXiv, version 2] with $\mu(S) = \eta(S) = |S|/n$ and $H = \rho$. Theorem 2.1, as is, does not deal with weights $w_u$, so we need to very slightly change the algorithm and proof. We add an extra SDP constraint

$$\sum_{u \in V} \|\bar{u}\|^2 w_u \leq W.$$

This constraint is clearly satisfied in the integral solution. We also change the function $f'$ (see page 10 of [8]). We let (here $w_u$ are the weights of vertices)

$$f''(S) = f'(S) - \underbrace{\frac{w(S)}{32W} \times H}_{\text{new term}}$$

$$\equiv \eta(S) - \frac{\delta(S)}{|E|} \times \frac{H}{4D \times SDP}$$

$$- \frac{\mu(S)}{4\rho} \times H - \frac{w(S)}{32W} \times H$$

$$= \frac{3|S|}{4n} - \frac{E(S, V \setminus S)}{|E|} \times \frac{\rho}{4D \times SDP} - \frac{w(S)}{32W} \times \rho.$$

Since,

$$\mathbb{E} \frac{w(S)}{32W} \times H \leq \sum_{u \in V} \frac{w_u \alpha \|\bar{u}\|^2}{32W} \times H \leq \sum_{u \in V} \frac{\alpha W}{32W} \times H = \frac{\alpha H}{32},$$

we get (compare with the third formula on page 10 [8]),

$$\mathbb{E}[f]''(S) \geq \mathbb{E}[f]'(S) - \frac{\alpha H}{32} \geq \frac{\alpha H}{32}.$$

This is sufficient for analysis in [8]. The SSE algorithm finds a set $S$ with $f''(S) > 0$. This condition implies that

$$\frac{3|S|}{4n} > \frac{w(S)}{W} \times \rho$$

and, consequently (as $|S| = \Theta(\rho n)$), $w(S) \leq O(W)$.

## C  SDP relaxations

**Minimum Balanced Cut:** The input to the problem is a graph $G(V, E)$, and the objective is to find a set $S$ of size $n/2$ with the minimum number of edges crossing it.

$$\min \quad \frac{1}{4} \sum_{(u,v) \in E(G)} \|\bar{u} - \bar{v}\|^2$$

subject to

$$\frac{1}{4} \sum_{u,v \in V} \|\bar{u} - \bar{v}\|^2 \geq \frac{n^2}{2} \quad \text{(Spreading constraint)}$$

$$\text{for all } u, v, w \in V, \quad \|\bar{u} - \bar{v}\|^2 + \|\bar{v} - \bar{w}\|^2 \geq \|\bar{u} - \bar{w}\|^2 \quad (\ell_2^2\text{-triangle inequalities})$$

$$\text{for all } u \in V, \quad \|\bar{u}\|^2 = 1$$

**Crude SDP (C-SDP) for Small-Set Expansion (SSE):** The input to the problem is a graph $G(V, E)$ and a parameter $\rho$, and the objective is to find a set $S$ of size $\rho n$ with the smallest number of edges crossing it.

$$\min \quad \frac{1}{2} \sum_{(u,v) \in E(G)} \|\bar{u} - \bar{v}\|^2$$

subject to

$$\text{for all } u \in V, \quad \sum_{v \in V} \langle \bar{u}, \bar{v} \rangle \leq \rho n \quad \text{(Spreading constraints)}$$

$$\text{for all } u, v, w \in V, \quad \|\bar{u} - \bar{v}\|^2 + \|\bar{v} - \bar{w}\|^2 \geq \|\bar{u} - \bar{w}\|^2 \quad (\ell_2^2\text{-triangle inequalities})$$

$$\text{for all } u, v \in V \quad \langle \bar{u}, \bar{v} \rangle \geq 0$$

$$\text{for all } u \in V, \quad \|\bar{u}\|^2 = 1$$

**Minimum Multicut:** The input to the problem is a graph $G(V, E)$ and a set of $k$ source-sink pairs $\{(s_i, t_i)\}_{1 \leq i \leq k}$, and the objective is to find a partition $\mathcal{P}'$ of the graph with minimum number of edges across partitions such that for all $i$, $\mathcal{P}'(s_i) \neq \mathcal{P}'(t_i)$.

$$\min \quad \frac{1}{2} \sum_{(u,v) \in E(G)} \|\bar{u} - \bar{v}\|^2$$

subject to

$$
\begin{aligned}
&\text{for all } 1 \leq i \leq k, && \langle \bar{s}_i, \bar{t}_i \rangle = 0 \\
&\text{for all } u, v, w \in V, && \|\bar{u} - \bar{v}\|^2 + \|\bar{v} - \bar{w}\|^2 \geq \|\bar{u} - \bar{w}\|^2 && (\ell_2^2\text{–triangle inequalities}) \\
&\text{for all } u \in V, && \|\bar{u}\|^2 = 1
\end{aligned}
$$